\documentclass[10pt,conference,letterpaper]{IEEEtran}
\usepackage[USenglish]{babel}
\usepackage{times,amsmath}
\usepackage[pdftex]{graphicx}
   \graphicspath{{./pdf/}{./png/}{./jpeg/}{./eps/}{./pst/}}
   \DeclareGraphicsExtensions{.pdf,.png,.jpeg,.eps}
\usepackage{cite}
\usepackage{url}
\usepackage{fmtcount}
\usepackage{color}
\usepackage[tight,footnotesize]{subfigure}
\usepackage{multicol}
\usepackage{amssymb}
\usepackage{wasysym}
\usepackage{nidanfloat} 
\usepackage{calc}
\usepackage{fix-cm}
\usepackage[activate={true,nocompatibility},final,tracking=true,factor=500,stretch=40,shrink=40]{microtype}
\usepackage{alltt}
\def\cal{\fam2}
\renewcommand{\d}{{\mathrm d}}

\newcommand{\sgn}{{\mathrm{sgn}}}

\newcommand{\Seps}{{{\cal{S}}_{\scriptscriptstyle \varepsilon}}}
\definecolor{DarkRed}{rgb}{0.5,0,0}
\definecolor{DarkGreen}{rgb}{0,0.5,0}
\definecolor{DarkerGreen}{rgb}{0,0.3333,0}
\definecolor{DarkBlue}{rgb}{0,0,0.75}
\definecolor{RoyalBlue}{rgb}{0,0.1373,0.4000}
\definecolor{NavyBlue}{rgb}{0,0,0.5020}
\definecolor{CobaltBlue}{rgb}{0,0.2784,0.6706}
\definecolor{lightlightgray}{rgb}{0.96875,0.96875,0.96875}
\definecolor{cyan}{rgb}{0,1,1}
\newcommand{\beginlabel}[2]{%
\begin{#1}\label{#2}}
\begin{document}
\pagestyle{plain}
\title{Quantile Tracking Filters for Robust Fencing\\ in Intermittently Nonlinear Filtering}
\author{\IEEEauthorblockN{Alexei V. Nikitin}
\IEEEauthorblockA{
Nonlinear LLC\\
Wamego, Kansas, USA\\
E-mail: avn@nonlinearcorp.com}
\and
\IEEEauthorblockN{Ruslan L. Davidchack}
\IEEEauthorblockA{Dept. of Mathematics, U. of Leicester\\
Leicester, UK\\
E-mail: rld8@leicester.ac.uk}}
\maketitle
\thispagestyle{headings}
\begin{abstract}
Robust fencing is an essential component of intermittently nonlinear filtering for mitigation of outlier interference. In such filtering, the upper and the lower fences establish a robust range that excludes noise outliers while including the signal of interest and the non-outlier noise. Then, the outlier values are replaced with those in mid-range. To increase the effectiveness of outlier noise identification, and to minimize the false negatives, the fences need to be both {\em tight\/} and {\em robust to outlier noise\/}. On the other hand, to minimize the false positives and to avoid the detrimental effects, such as instabilities and excessive distortions, often associated with nonlinear filtering, the fences need to be {\em inclusive\/}, so that the signal of interest and the non-outlier noise remain within the fences. Quantile Tracking Filters (QTFs) are an appealing choice for such robust fencing in intermittently nonlinear filtering, as QTFs are {\em analog\/} filters suitable for wideband real-time processing of continuous-time signals and are easily implemented in analog circuitry. Further, their numerical computations are {\boldmath$\mathcal{O}(1)$} per output value in both time and storage, which also enables their high-rate digital implementations in real time. In this paper, we first provide a brief general discussion of the outlier noise and its mitigation by intermittently nonlinear filters. We then focus on the basic properties of the QTF-based fencing, discuss the approaches to choosing its parameters, and illustrate the use of the QTF fencing for various types of the signal+noise mixtures. We also demonstrate how the {\em Complementary\/} Intermittently Nonlinear Filtering (CINF) arrangements allow us to increase the tightness and robustness of the QTF fencing, while preserving its inclusivity, and to enable the mitigation of outlier noise obscured by high-amplitude non-outlier signals such as the signal of interest itself.
\end{abstract}
\maketitle

\begin{figure}[!b]
\centering{\includegraphics[width=8.6cm]{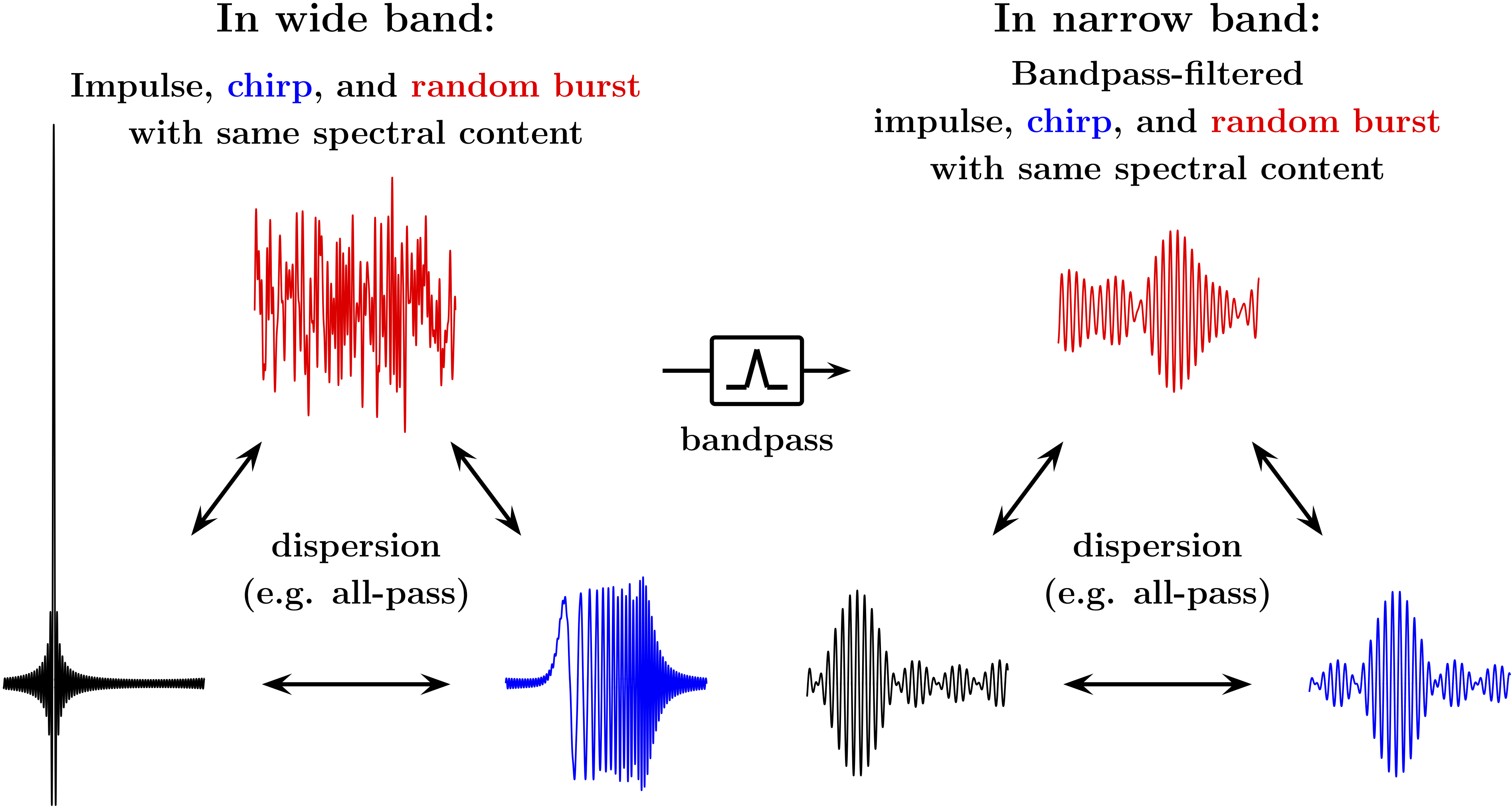}}
\caption{Effect of filtering on temporal and amplitude structure of signal is more apparent at wider bandwidth.
\label{fig:imp2chirp}}
\end{figure}

\section{Outlier noise and its mitigation by intermittently nonlinear filtering} \label{sec:introduction}
In addition to ever-present thermal noise, various communication and sensor systems can be affected by interfering signals that originate from a multitude of other natural and technogenic (man-made) phenomena. Such interfering signals often have intrinsic temporal and/or amplitude structures different from the Gaussian structure of the thermal noise.  Specifically, when observed in the time domain, the interference may contain distinct amplitude outliers. The presence of different types of such outlier noise is widely acknowledged in multiple applications, under various general and application-specific names, most commonly as {\it impulsive\/}, {\it transient\/}, {\it burst\/}, or {\it crackling\/} noise.

For example, the outlier interference can be produced by some ``countable" or ``discrete" relatively short duration events that are separated by relatively long periods of inactivity. Provided that the observation bandwidth is sufficiently large relative to the rate of these non-thermal noise generating events, and depending on the noise coupling mechanisms and the system's filtering properties and propagation conditions, such noise may contain distinct  transients that appear as time-domain outliers. More generally, apparent outliers in a signal can appear, disappear, and reappear due to various filtering effects, including dispersion, fading and multipath, as the signal propagates through media and/or the signal processing chain. In the analog domain, such filtering can be viewed as a linear combination of the signal with its derivatives and antiderivatives (e.g. convolution) of various orders. In the digital domain, it is a combination of differencing and summation operations.

Typically, the effect of filtering on the temporal and/or the amplitude structure of a signal would be more apparent at wider bandwidths, as broadening the bandwidth increases the time resolution. This is illustrated in Fig.~\ref{fig:imp2chirp}, where the impulse, the chirp, and the random burst signals have the same spectral contents, and only the phases in their Fourier representations are different. Thus these three signals can be morphed into each other by all-pass filtering that leaves their power spectral densities (PSDs) unmodified. In a wide band, such filtering drastically changes the time-domain appearance of the signal and its amplitude density, while in a narrow band (after the bandpass filtering) these changes are much less apparent and all signals maintain similar temporal and amplitude structures.

\begin{figure}[!t]
\centering{\includegraphics[width=8.6cm]{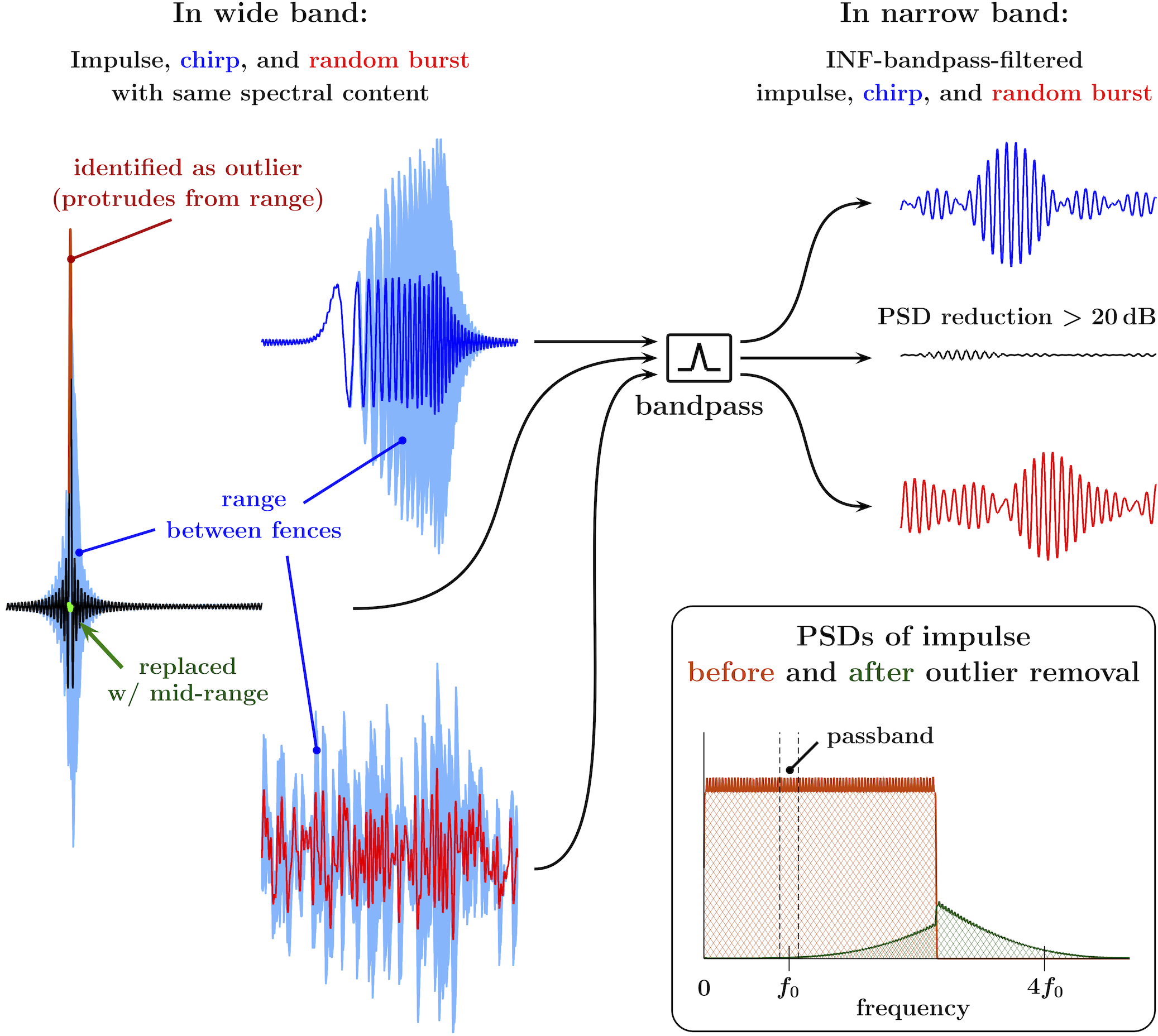}}
\caption{INF significantly affects impulse signal, leaving chirp and random burst signals unchanged.
\label{fig:imp2chirp INF}}
\end{figure}
\begin{figure*}[!b]
\centering{\includegraphics[width=16cm]{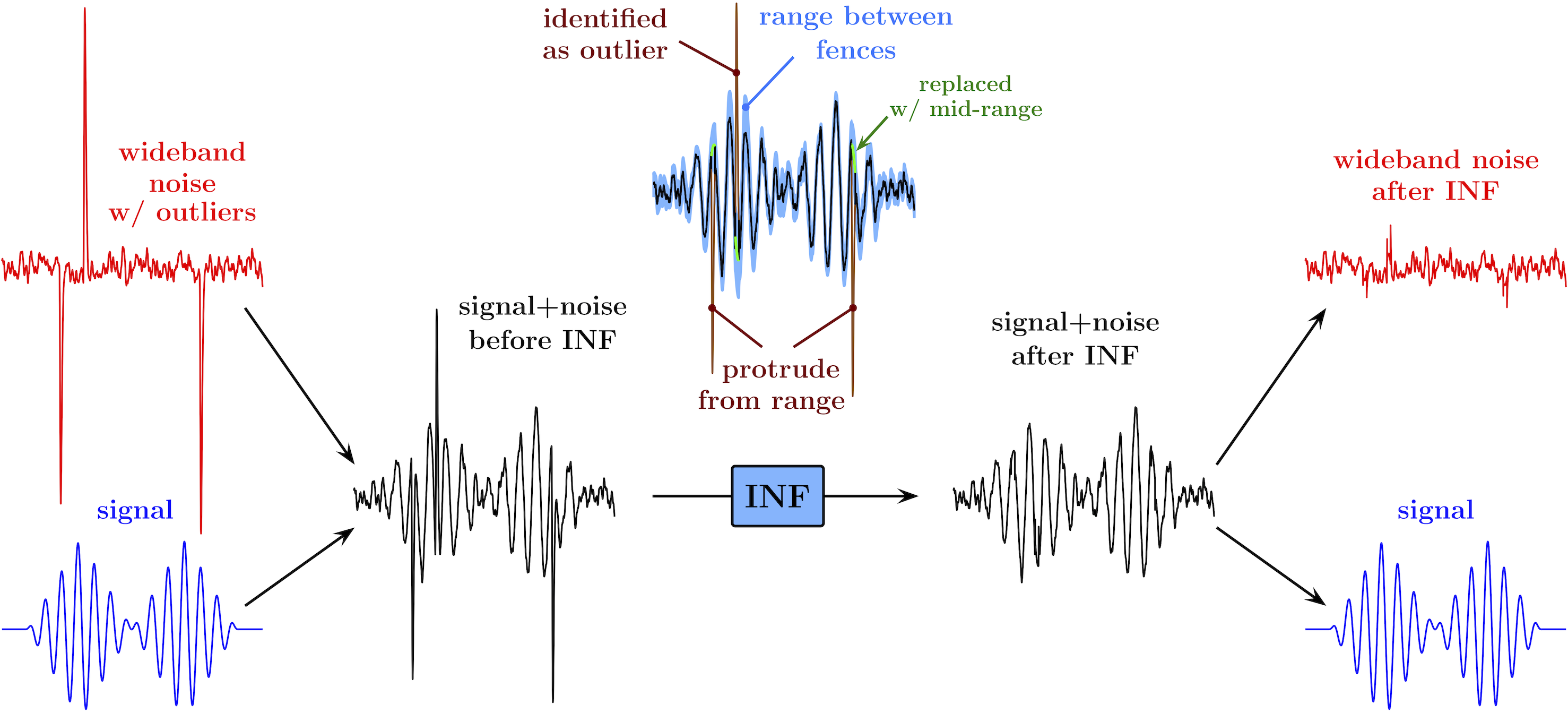}}
\caption{{\bf Intermittently Nonlinear Filtering (INF) removes outliers from wideband noise:}
Noise outliers are identified as protrusions outside of fenced range around signal, and their values are replaced by those in mid-range.
\label{fig:INF fencing}}
\end{figure*}
\begin{figure*}[!b]
\centering{\includegraphics[width=17.6cm]{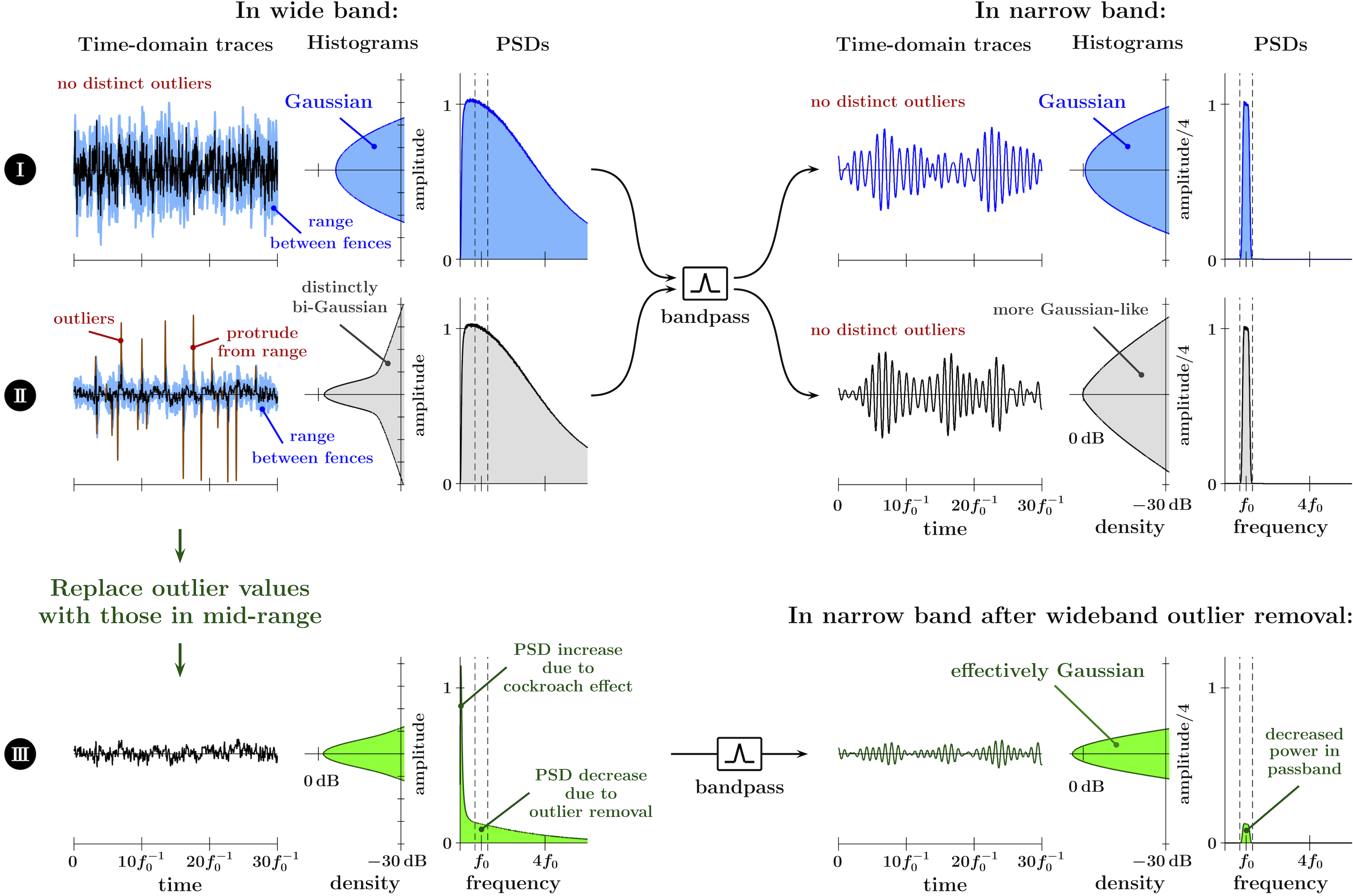}}
\caption{{\bf Illustrative example of effect of outlier removal from wideband noise on noise PSD:}
Mitigation of outliers in wideband noise reduces its PSD in band of interest. Increased PSD at low frequencies is due to so-called "cockroach effect"~\cite{Nikitin19hidden}.
\label{fig:INF noise}}
\end{figure*}

\subsection{Intermittently Nonlinear Filtering (INF)} \label{subsec:INF}
Nonlinear filters are capable of disproportionately affecting signals with distinct temporal and/or amplitude structures, even when these signals have identical PSDs, overall or in the band of interest. Thus nonlinear filtering enables in-band mitigation of interfering signals with the temporal and/or amplitude structures sufficiently different from that of the signal of interest to levels unattainable by linear filters~\cite{Nikitin12aHSDPA, Nikitin13adaptive, Nikitin15OOB}. Yet this advantage can often be negated by the detrimental effects, such as distortions and instabilities, often associated with nonlinear filtering in general. However, when the main goal is the mitigation of outlier interference, the detrimental effects of nonlinear filtering can be significantly reduced by allowing a filter to behave linearly when the input signal does not contain outliers, and respond nonlinearly only when the outliers are encountered. Since high-amplitude outliers make disproportionately large contribution to the overall power of the interference, such {\em Intermittently Nonlinear\/} Filtering (INF) allows us to significantly improve the signal quality when the interference contains outliers, while mainly preserving linear, ``no harm" behavior otherwise.

The basic concept of a particular type of INF for mitigation of such outlier interference, introduced in~\cite{Nikitin18ADiC-ICC, Nikitin19hidden, Nikitin19complementary, Nikitin19ADiCpatentCIP1}, can be briefly described as follows: First, we establish a robust {\em range\/} around the signal, which excludes noise outliers while including the signal of interest and the non-outlier noise; then, we replace the signal values that extend outside of the range with those within the range.

Let us first illustrate the application of such filtering to the wideband impulse, chirp, and random burst signals with the same spectral content, where the range is established by the Quantile Tracking Filters (QTFs) as described further in this paper. As shown in Fig.~\ref{fig:imp2chirp INF}, because the chirp and the random burst signals contain no distinct outliers that protrude from the range, an INF with such fencing will be fully transparent to the chirp and the random burst. In contrast, the wideband impulse contains a distinct outlier, and, as indicated by the orange line fragment in the impulse, this outlier is identified as a protrusion from the range. While having a relatively small duration, this outlier contains most of the total power of the impulse, and, as shown in the lower right of Fig.~\ref{fig:imp2chirp INF}, replacing the outlier values by those in mid-range (shown by the green line fragment in the impulse) significantly affects the resulting PSD. Thus, in this example, the power in the passband around~$f_0$ (after the bandpass filer) is reduced by over~$20\,$dB.

Also note that even if a similar short-duration portion of either the chirp or the random burst signal were misidentified as an outlier and replaced by the respective mid-range values, this would have a significantly smaller effect on the resulting narrowband signal than the removal of the outlier in the impulse signal. Since both the chirp and the random burst have much smaller peak-to-average power ratios (PARs) compared to the impulse, such a small portion of a non-outlier signal does not make a disproportionally large contribution to its overall PSD. 

Fig.~\ref{fig:INF fencing} illustrates the basic concept of the INF introduced in~\cite{Nikitin18ADiC-ICC, Nikitin19hidden, Nikitin19complementary, Nikitin19ADiCpatentCIP1} as it applies to removal of outliers from the wideband noise affecting a narrower-band signal of interest. The main challenge is in constructing the fences that are sufficiently ``tight" and robust to noise outliers yet ``inclusive," so that the range between them fully contains the signal of interest and the non-outlier noise. With such fencing, the outlier values are replaced by those that are significantly closer to the values of the signal of interest, and the resulting INF output can be viewed as the signal of interest affected by a wideband noise with reduced outlier component.

We would like to emphasize at this point that the purpose of the INF is to remove the wideband noise outliers first, before any subsequent filtering is performed. In this regard, INF is just a {\em pre-filtering\/} to remove wideband noise outliers, and it needs to be followed by whatever linear filtering would be used otherwise. If such outliers are present, the resulting signal quality can be increased beyond the level attainable by linear filtering alone. Otherwise, the INF does not affect the signal+noise mixture, which results in the overall linear, ``no harm" behavior.

Fig.~\ref{fig:INF noise} provides an illustrative example of the effect of such INF-based outlier removal from wideband noise on the resulting noise PSD. First note that the wideband noise shown in the left-hand side of row~I is Gaussian, it does not contain distinct outliers protruding from the range, and it may be considered to be effectively outlier-free. Since no outliers are removed, the effect of INF is simply that of ``no harm" and the noise PSD is not affected.
In contrast, the noise shown in the left-hand side of row~II, while having the same PSD as the noise in row~I, is impulsive and contains distinct amplitude outliers. Without outlier removal, the result of filtering of this wideband impulsive noise to within the band of interest (around~$f_0$) is effectively equivalent to that of filtering the Gaussian noise shown in row~I. However, as shown in row~III, mitigation of the outliers in the wideband noise of row~II significantly (by about~10\,dB) reduces its PSD in the band of interest. Thus INF improves the signal quality when the wideband noise contains an outlier component, causing no harm otherwise. Note that, after the outlier removal, the increase in the noise PSD at low frequencies is due to the so-called "cockroach effect" described in~\cite{Nikitin19hidden}.

As discussed in~\cite{Nikitin19hidden, Nikitin19complementary}, when we are not constrained by the needs for either analog or wideband, high-rate real-time digital processing, in the digital domain such INF function can perhaps be accomplished by a {\it Hampel filter\/}~\cite{Hampel74influence} or by one of its variants~\cite{Pearson16generalized}. In a Hampel filter the ``mid-range" is calculated as a windowed median of the input, and the range is determined as a scaled absolute deviation about this windowed median. However, the windowed median estimation in a Hampel filter relies on the operation of sorting, and its analog implementation meets with considerable conceptual and practical difficulties~\cite{Nikitin03signal, Nikitin04adaptive}. Further, in order to be robust to outliers with the typical width~$\Delta{T}$, the width~$T$ of the window for the median filter needs to be sufficiently larger than~$2\Delta{T}$. Thus, for a sampling rate~$F_{\rm s}$, numerical computations of a windowed median require $\mathcal{O}\left(TF_{\rm s}\log(TF_{\rm s})\right)$ per output value in time, and $\mathcal{O}(TF_{\rm s})$ in storage, becoming prohibitively expensive for high-rate real-time processing.

\section{Robust real-time fencing for INF} \label{sec:robust fencing}
\begin{figure}[!t]
\centering{\includegraphics[width=8.6cm]{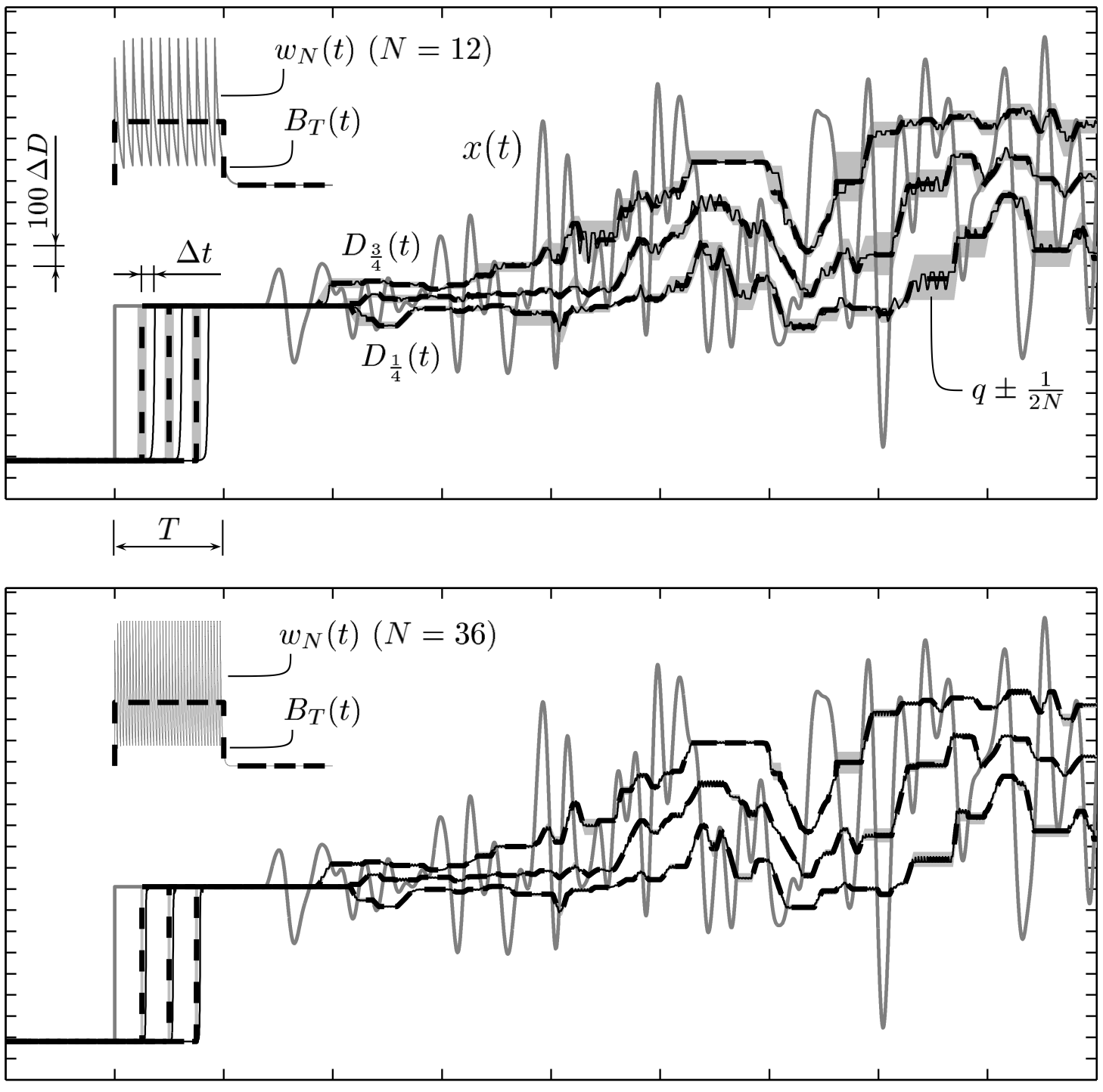}}
\caption{Illustration of performance of analog rank filter given by Eq.~(9) in~\cite{Nikitin04adaptive} by comparing its quartile outputs $D_q(t)$ (for $q = 1/4$, $1/2$, and $3/4$, solid black lines) with respective outputs of ``exact" order statistic filters in rectangular moving window~$B_T(t)$ of width~$T$ (dashed lines).
(Reproduced from~\cite{Nikitin04adaptive}.)
\label{fig:adaptive approximation}}
\end{figure}

While in a Hampel filter the range is established around a windowed median of the input, a robust range ${[\alpha_-,\alpha_+]}$ that excludes outliers of a signal can also be formed around a different robust measure of the central tendency, for example, the {\em midhinge\/}~\cite{Tukey77exploratory}.
Hence it can be obtained as a range between {\it Tukey's fences\/}~\cite{Tukey77exploratory} constructed as linear combinations of the 1st $\left(Q_{[1]}\right)$ and the 3rd $\left(Q_{[3]}\right)$ quartiles of the signal in a moving time window:
\beginlabel{equation}{eq:Tukey's fences}
  [\alpha_-,\alpha_+] = {\big [}Q_{[1]}\!-\!\beta\left(Q_{[3]}\!-\!Q_{[1]}\right)\!,\,Q_{[3]}\!+\!\beta\left(Q_{[3]}\!-\!Q_{[1]}\right)\!{\big ]},
\end{equation}
$\!$where $\alpha_+$, $\alpha_-$, $Q_{[1]}$, and $Q_{[3]}$ are time-varying quantities, and $\beta$ is a positive scaling parameter of order unity (e.g. $\beta=1.5$).  To enable analog and/or real-time digital (e.g., with~$\mathcal{O}(1)$ per output value in both time and storage) realizations of such fencing, one can employ approximations for the quartile values~$Q_{[1]}(t)$ and~$Q_{[3]}(t)$ in a moving window of time, that are suitable for implementation in analog feedback circuits. For example, the adaptive approximation described in~\cite{Nikitin04adaptive} approximates a boxcar moving window ${B_T(t) = \left[ \theta(t) - \theta(t\!-\!T) \right]/T}$, where~$\theta(x)$ is the Heaviside unit step function, by the window~$w_N(t)$ consisting of~$N$ exponential kernels:
\beginlabel{equation}{eq:boxcar time response}
  w_N(t) = \frac {1}{N} \sum_{k=0}^{N-1} h_{\tau}(t-2k\tau)\,,
\end{equation}
where $\tau = T/(2N)$ and~${h_{\tau}(t) = \theta(t)\, \exp \left( -\frac {t}{\tau}\!-\!\ln\tau \right)}$. For such a window, a rank filter can be approximately expressed as a system of 1st~order differential equations, and hence realized in an analog feedback circuit. The accuracy of this approximation improves with the increase in the number~$N$ of the kernels in the approximation, as the timing (delay) error~$\Delta{t}$ is inversely proportional to~$N$, and the residual oscillations of the approximate outputs occur within the ${q \pm 1/(2N)}$ intervals around the respective outputs of the ``exact" quantile filters. This is illustrated in Fig.~\ref{fig:adaptive approximation} (reproduced from~\cite{Nikitin04adaptive}).

While conceptually this approximation enables analog and/or real-time digital implementations of rank filters, a large number~$N$ of the kernels required for its adequate accuracy (e.g. $N\gtrsim 10$) significantly increases the computational burden and the memory requirements, and effectively prohibits its practical hardware development. Favorably, robust fencing for effective use in INF can be achieved by much simpler means, e.g. by the {\em Quantile Tracking Filters\/} described further in this section.

\subsection{Quantile Tracking Filters (QTFs)} \label{subsec:QTFs}
\begin{figure}[!t]
\centering{\includegraphics[width=8.6cm]{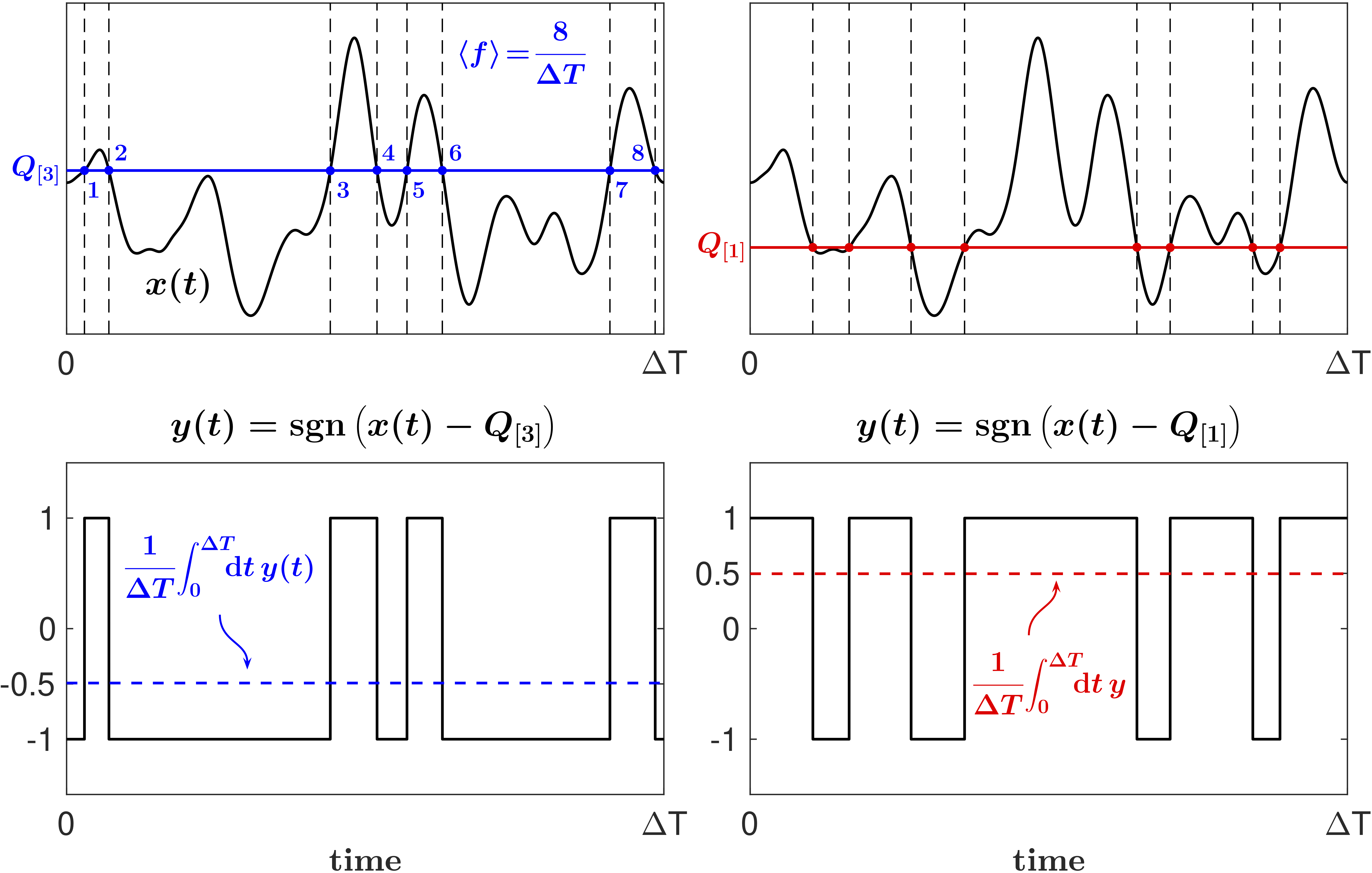}}
\caption{``Exact" quartiles of~$x(t)$ in interval~$[0,{\Delta{T}}]$ as solutions of~(\ref{eq:Qq exact}).
\label{fig:QTF ideal}}
\end{figure}

First note that the time-independent~$Q_q$ that satisfies the equality
\beginlabel{equation}{eq:Qq exact}
  \frac{1}{\Delta{T}}\!\int_0^{\Delta{T}}\!\!\!\d{t}\, \sgn\left(x(t)-Q_q\right) = 1\!-\!2q\,,
\end{equation}
where $0<q<1$, represents the $q$-th quantile of the signal~$x(t)$ in the time interval~$[0,{\Delta{T}}]$. Indeed, from~(\ref{eq:Qq exact}) follows that a~$q$-th fraction of~$x(t)$ in this interval has the values smaller than~$Q_q$. For example, as illustrated in Fig.~\ref{fig:QTF ideal}, for~$q=3/4$ the values of~$x(t)$ are smaller than~$Q_q$ for the three quarters of the total time interval~$[0,{\Delta{T}}]$.

\begin{figure}[!t]
\centering{\includegraphics[width=8cm]{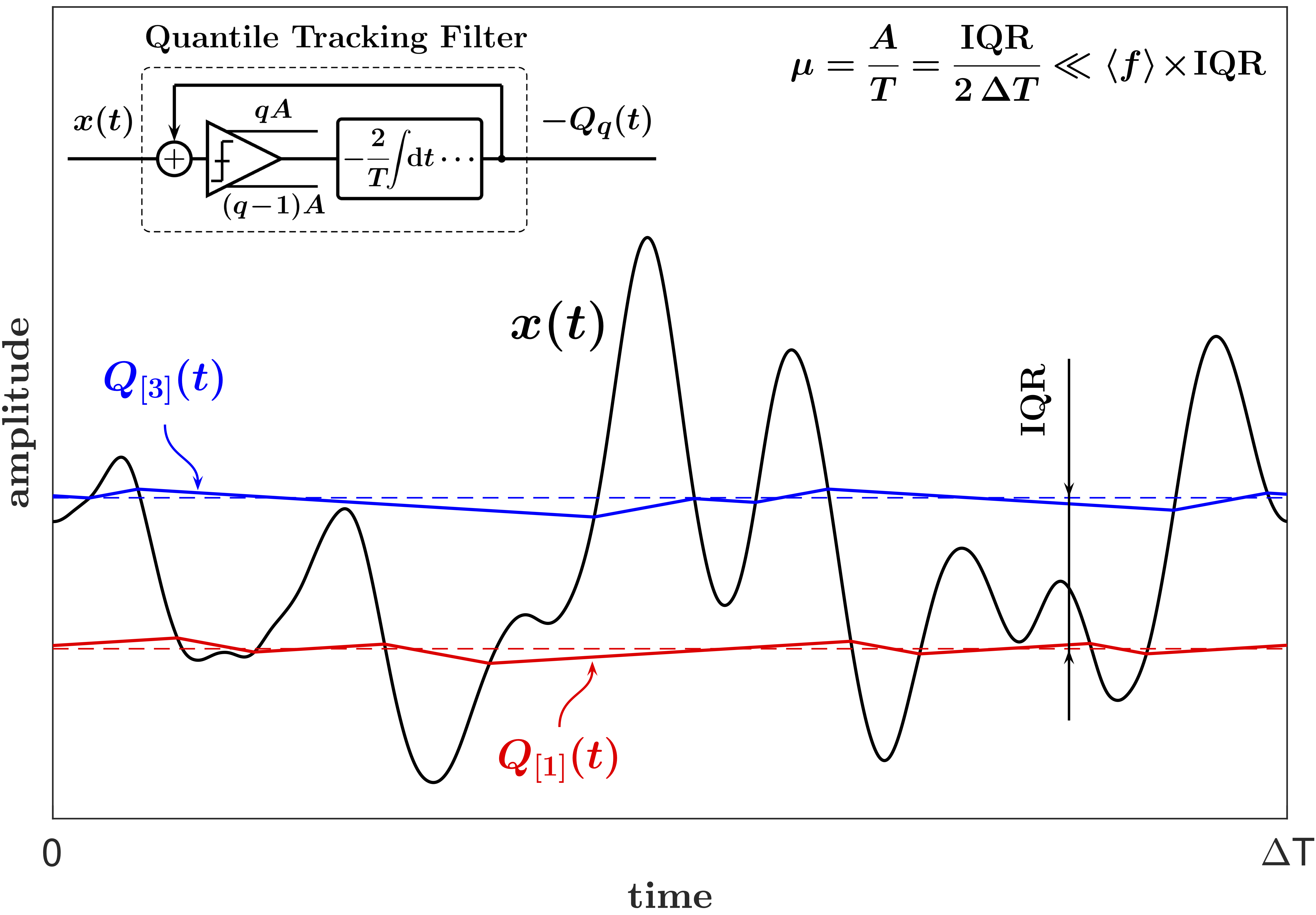}}
\caption{Outputs~$Q_{[3]}(t)$ and~$Q_{[1]}(t)$ of QTFs approximating ``exact" 3rd and 1st quartiles, respectively, of~$x(t)$ in moving time window.
\label{fig:QTF approx}}
\end{figure}

Now let $x(t)$ be a continuous stationary signal with a constant mean and a positive interquartile range (IQR), characterized by an average crossing rate~$\langle f\rangle$ of $x(t)$ with a (constant) threshold equal to the $q$-th quantile of~$x(t)$, and consider a time-dependent~$Q_q(t)$ given by the following differential equation:
\beginlabel{equation}{eq:Qq QTF}
  \dot{Q}_q = \mu\, \left[\sgn(x\!-\!Q_q) + 2q-1\right]\,,
\end{equation}
where~$\mu$ is the {\em rate parameter\/} and $0<q<1$ is the {\em quantile parameter\/}. Note that~(\ref{eq:Qq QTF}) can be solved by a simple analog circuit shown in the upper left corner of Fig.~\ref{fig:QTF approx}.

As follows from~(\ref{eq:Qq QTF}), for any~$t=t_i$ such that~$x(t_i)=Q_q(t_i)$ a {\em zero crossing\/} of~$x(t)\!-\!Q_q(t)$ will occur if $\dot{x}(t_i)>2q\mu$ (for an {\em upward\/} crossing) or $\dot{x}(t_i)<2(q\!-\!1)\mu$ (for a {\em downward\/} crossing). Thus, for a sufficiently small~$\mu$ (e.g., much smaller than the product of the IQR and~$\langle f\rangle$), $Q_q(t)$ will be a piecewise-linear signal consisting of alternating segments with a positive slope~$2q\mu$ when~$x(t)>Q_q(t)$ and a negative slope~$2(q\!-\!1)\mu$ when~$x(t)<Q_q(t)$. Further, an approximate stationary solution of~(\ref{eq:Qq QTF}) can be written implicitly as
\beginlabel{equation}{eq:Qq steady}
  \overline{\sgn\left(x(t)\!-\!Q_q(t)\right)} \approx 1\!-\!2q\,,
\end{equation}
where the overline denotes averaging over some time interval~${\Delta{T}\gg \langle f\rangle^{-1}}$. Therefore, as illustrated in Fig.~\ref{fig:QTF approx}, for these conditions $Q_q(t)$ approximates the $q$-th quantile of $x(t)$ in the time interval~$\Delta{T}$.

Since outputs of analog QTFs are piecewise-linear signals consisting of alternating segments with positive and negative slopes, care should be taken in finite difference implementations of QTFs to avoid the ``overshoots" around the crossings of~$Q_q(t)$ with~$x(t)$. In particular, when $x(n)\!-\!Q_q(n\!-\!1)$ is outside of the interval ${h\mu\left[2(q\!-\!1),2q\right]}$, where~$h$ is the time step, one may set $Q_q(n)\!=\!x(n)$, as illustrated in~\cite{Nikitin19ADiCpatentCIP1} and in Appendix~\ref{app:numerical}.

\subsubsection{QTF parameters for signal with linear trend} \label{subsubsec:linear trend}
\begin{figure}[!b]
\centering{\includegraphics[width=8.6cm]{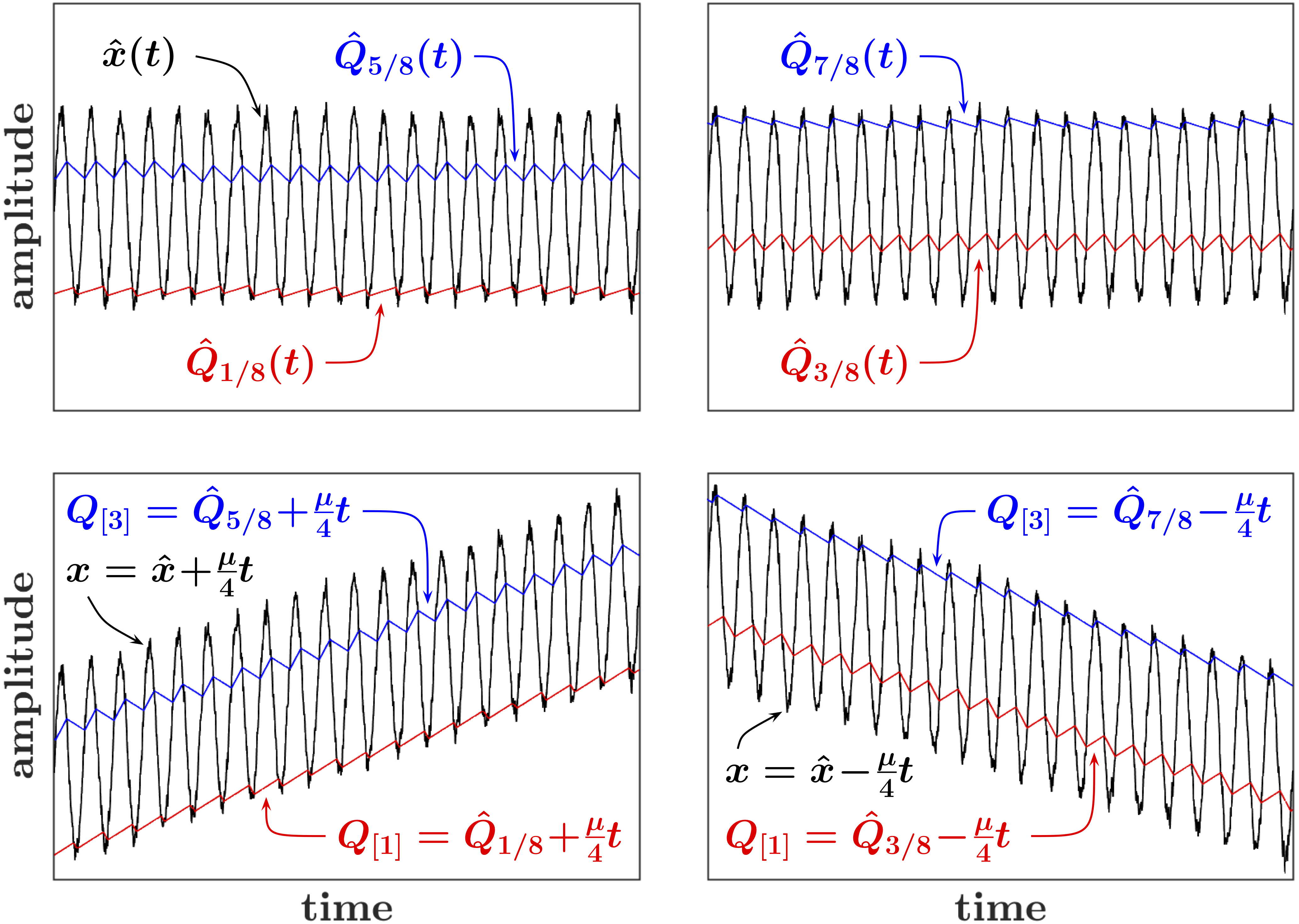}}
\caption{QTF outputs for signal with linear trend.
\label{fig:trend}}
\end{figure}

For a signal~$x(t)$ with a linear trend,
\beginlabel{equation}{eq:x with trend}
  x(t) = \hat{x}(t) \pm \hat{\mu}(t\!-\!t_0),
\end{equation}
the solution~$Q_q(t)$ of~(\ref{eq:Qq QTF}) can be represented as
\beginlabel{equation}{eq:Q_q with trend}
  Q_q(t) = \hat{Q}_{\hat{q}}(t) \pm \hat{\mu}(t\!-\!t_0),
\end{equation}
where
\beginlabel{equation}{eq:Qq prime}
  \dot{\hat{Q}}_{\hat{q}} = \mu\, \left[\sgn(\hat{x}\!-\!\hat{Q}_{\hat{q}}) + 2\hat{q}-1\right],
\end{equation}
and where~${\hat{q}=q\mp \hat{\mu}/(2\mu)}$. For example, as illustrated in Fig.~\ref{fig:trend} for ${x(t) = \hat{x}(t) \pm \mu t/4}$, ${\hat{q}=(6\mp 1)/8}$ for ${q=3/4}$ and~${\hat{q}=(2\mp 1)/8}$ for ${q=1/4}$. Therefore, to accommodate such a linear trend while satisfying the condition~$0<\hat{q}<1$, for a given quantile parameter~$q$ the QTF rate parameter~$\mu$ must satisfy the inequality 
\beginlabel{equation}{eq:mu4trend}
  \mu > \frac{\hat{\mu}}{2\min(q,1\!-\!q)}.
\end{equation}

\begin{figure*}[!b]
\centering{\includegraphics[width=16cm]{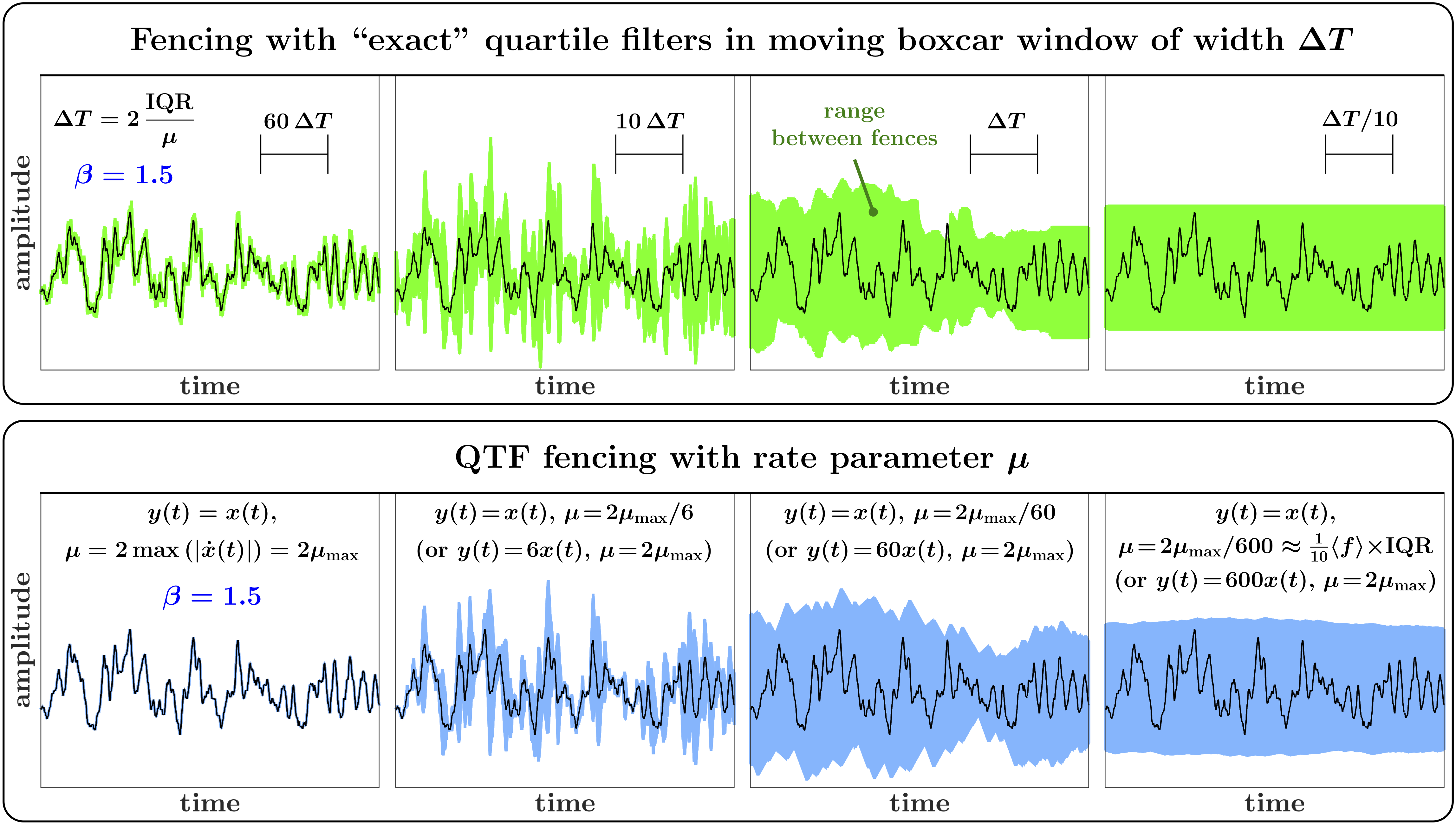}}
\caption{Overall behavior of QTF fencing is similar to that with ``exact" quartile filters in moving boxcar window of width~$\Delta{T}=2\times{\rm IQR}/\mu$.
\label{fig:RANKvsQTFs4noise}}
\end{figure*}
\begin{figure*}[!t]
\centering{\includegraphics[width=16.4cm]{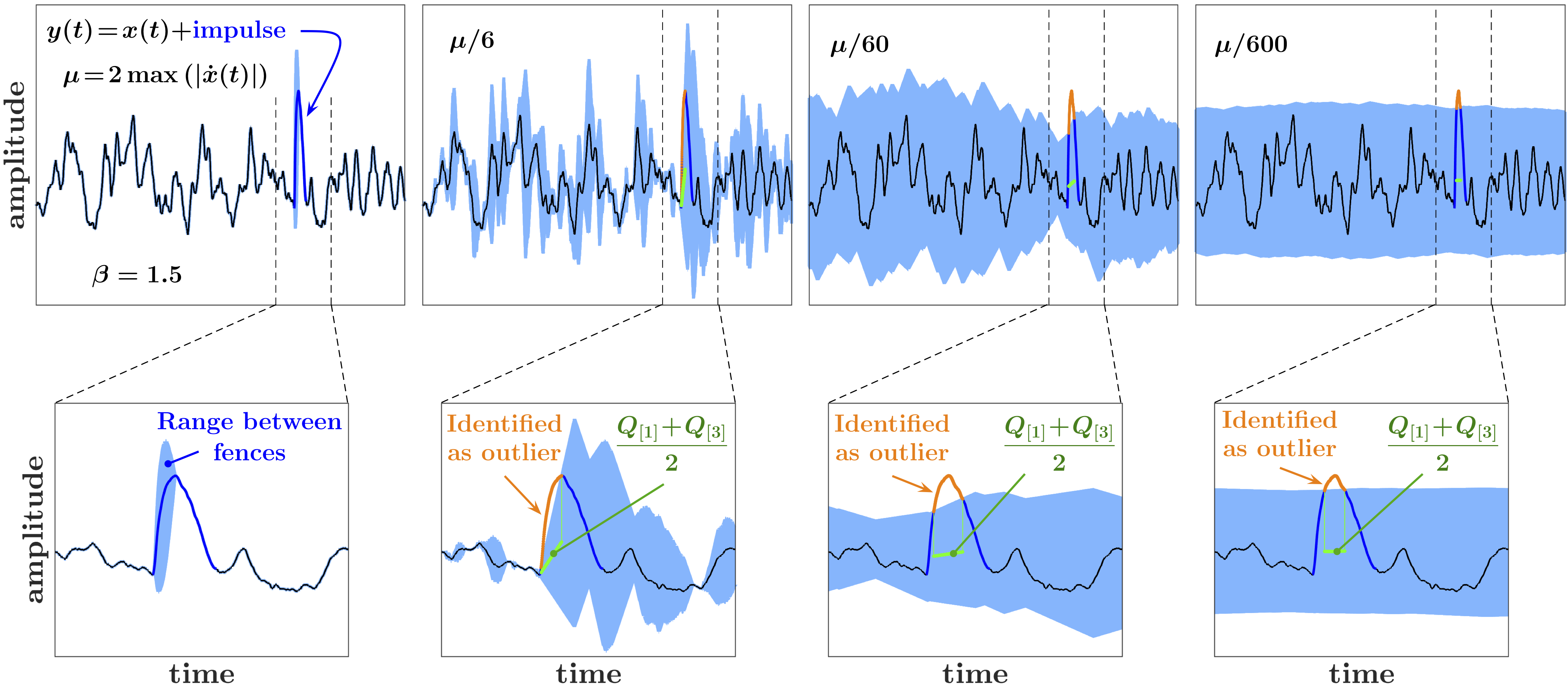}}
\caption{QTF fences with $\beta=1.5$ and different rate parameters for stationary random signal with added impulse.
\label{fig:QTFs4noise}}
\end{figure*}
\begin{figure*}[!b]
\centering{\includegraphics[width=16.4cm]{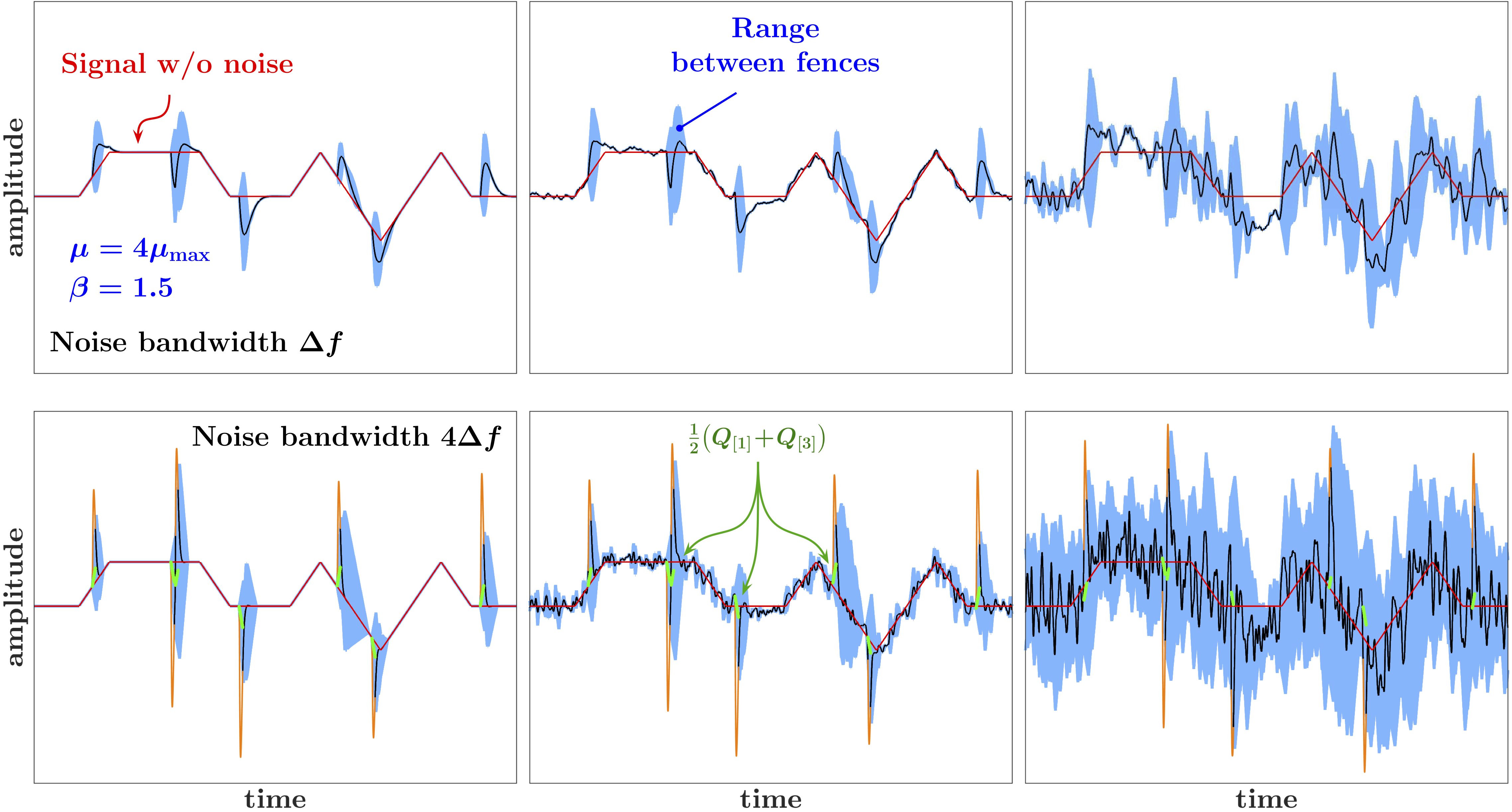}}
\caption{QTF fencing is more effective for identification and mitigation of outliers in wideband noise affecting slowly varying signal at higher observation bandwidths.
{\em Orange lines: Signal+noise protruding outside of QTF fences.}
\label{fig:QTFs4ramp}}
\end{figure*}

\subsection{QTF fences for robust range} \label{subsec:range}
Let us replace $\sgn(x)$ in~(\ref{eq:Qq QTF}) with the {\em comparator function\/}~$\Seps(x)$ such that $\lim_{\varepsilon\to 0}\Seps(x) = \sgn(x)\,$:
\beginlabel{equation}{eq:QTF eps}
  \dot{Q}_q = \mu\, \left[\lim_{\varepsilon\to 0}\Seps(x\!-\!Q_q) + 2q-1\right].
\end{equation}
In~(\ref{eq:QTF eps}), the comparator function can be any {\em continuous\/} function such that~$\Seps(x)=\sgn(x)$ for~$|x|\gg\varepsilon$, and $\Seps(x)$~changes monotonically from~``$-1$" to~``$1$" so that most of this change occurs over the range~$[-\varepsilon,\varepsilon]$. 

For example, for simplicity, we can express~$\Seps(x)$ as
\beginlabel{equation}{eq:comparator}
  \Seps(x) = \left\{
  \begin{array}{cc}
    \frac {x}{\varepsilon} & \mbox{for} \quad |x|<\varepsilon\\
    \sgn(x) & \mbox{otherwise}
  \end{array}\right.,
\end{equation}
where~$\varepsilon>0$. Then
\beginlabel{equation}{eq:comparator offset}
  \Seps(x)\!+\!2q\!-\!1 = \left\{
  \begin{array}{cc}
    \frac {x+(2q\!-\!1)\varepsilon}{\varepsilon} & \mbox{for} \quad |x|<\varepsilon\\
    \sgn(x)\!+\!2q\!-\!1 & \mbox{otherwise}
  \end{array}\right.,
\end{equation}
and $2(q\!-\!1)\le\Seps(x)\!+\!2q\!-\!1\le2q$. From this point on, we will always assume that the sign function~$\sgn(x)$ in~(\ref{eq:Qq QTF}) is represented as~$\lim_{\varepsilon\to 0}\Seps(x)$.

\subsubsection{Tightest possible fences for continuous signal} \label{subsubsec:tightest}
Let us now consider a continuous signal~$x(t)$ with a bounded amplitude of its first time derivative: ${\left|\dot{x}(t)\right| \le\mu_{\rm max}}$ for all~$t$.
For a sufficiently large~$\mu$,
\beginlabel{equation}{eq:mu}
  \mu > \mu_q = \frac{\mu_{\rm max}}{2\min(q,1\!-\!q)},
\end{equation}
for any initial condition~$Q_q\!=\!Q_q(t_0)$ in~(\ref{eq:QTF eps}) such that ${\left|x(t_0)-Q_q(t_0)\right|>\varepsilon}$, the inequality ${\left|x(t)-Q_q(t)\right|<\varepsilon}$ will be satisfied after a time interval~$\Delta{t}$ such that
\beginlabel{equation}{eq:Delta t}
  0\leq\Delta{t} \leq \frac{\left|x(t_0)-Q_q(t_0)\right|-\varepsilon}{\mu-\mu_q}.
\end{equation}
Then, for~${t\geq t_0\!+\!\Delta{t}}$ it follows from~(\ref{eq:QTF eps}) and~(\ref{eq:comparator offset}) that
\beginlabel{equation}{eq:QTF large mu}
  Q_q = x + \lim_{\varepsilon\to 0}\, \varepsilon \left[2q\!-\!1\!-\!\frac{1}{\mu}\dot{Q}_q \right] = x \quad {\rm for} \quad \mu > \mu_q,
\end{equation}
and~$Q_q(t)$ will be effectively equal to~$x(t)$. Further, the distance~$\Delta{\alpha}$ between the fences is
\beginlabel{equation}{eq:Tukey's range}
  \Delta{\alpha} = \alpha_+ - \alpha_- = (2\beta+1)\left(Q_{[3]} - Q_{[1]}\right),
\end{equation}
and for~${t\geq t_0\!+\!\Delta{t}}$
\beginlabel{equation}{eq:tightest range}
  \Delta{\alpha} = \lim_{\varepsilon\to 0} (2\beta+1)\varepsilon = 0  \quad {\rm for} \quad \mu \ge 2\mu_{\rm max}.
\end{equation}

Note that from~(\ref{eq:QTF large mu}) it also follows that, for a finite~$\varepsilon>0$ and~$\mu \ge \mu_q$, the output~$Q_q(t)$ of a QTF applied to the signal~$x(t)$ is equal to the output of a 1st~order lowpass filter with the corner frequency~$\mu/(2\pi\varepsilon)$ applied to the input signal~$x(t) + (2q\!-\!1)\varepsilon$.

\subsubsection{Less tight yet still inclusive fences} \label{subsubsec:less tight yet inclusive}
From now on, we will focus on the fences constructed, according to~(\ref{eq:Tukey's fences}), as linear combinations of the QTF outputs for $q=1/4$ $\left(Q_{[1]}(t)\right)$ and $q=3/4$ $\left(Q_{[3]}(t)\right)$.

In order for~$x(t)$ to protrude from the range~$\left[\alpha_-(t),\alpha_+(t)\right]$, $x(t)$ needs to cross~$\alpha_+(t)$ upward, or~$\alpha_-(t)$ downward at some point~$t_i$. As follows from~(\ref{eq:Qq QTF}) and~(\ref{eq:Tukey's fences}), this can only happen if
\beginlabel{equation}{eq:range protrusion}
  |\dot{x}(t_i)| > \left(\frac{3}{2}+\beta\right)\mu\,.
\end{equation}
Therefore, the range formed by the QTFs with the rate parameter~$\mu$ satisfying the inequality
\beginlabel{equation}{eq:sufficient mu}
  \mu \ge \frac{2\mu_{\rm max}}{3+2\beta},
\end{equation}
where~${\mu_{\rm max}=\max \left( \left|\dot{x}(t)\right| \right)}$, will fully contain~$x(t)$.

Note that~(\ref{eq:sufficient mu}) represents only a {\em sufficient\/} condition for the range~$\left[\alpha_-(t),\alpha_+(t)\right]$ to fully contain~$x(t)$, and~(\ref{eq:range protrusion}) quantifies the {\em robustness\/} of the fences to outliers:
In order for an outlier in~$x(t)$ to protrude from the range, not only its magnitude needs to be sufficiently large, but also its rising (for the upper fence) or falling (for the lower fence) rate of change at a fence needs to be sufficiently high. The smaller~$\mu$ (and/or $\beta$), the more robust are the fences.

\subsubsection{Overall behavior of QTF fencing} \label{subsubsec:overall}
As follows from the discussion in Section~\ref{subsec:QTFs}, for a continuous stationary signal with a constant mean and a positive IQR, the outputs~$Q_{[1]}(t)$ and~$Q_{[3]}(t)$ of QTFs with a sufficiently small rate parameter~$\mu$ will approximate the~1st and the~3rd quartiles, respectively, of the signal obtained in a moving boxcar time window with the width~$\Delta{T}$ of order~${2\times{\rm IQR}/\mu\gg\langle f\rangle^{-1}}$, where~$\langle f\rangle$ is the average crossing rate of~$x(t)$ with the~1st and the~3rd quartiles of~$x(t)$. Consequently, as illustrated in Fig.~\ref{fig:RANKvsQTFs4noise}, the overall behavior of the QTF fencing for a stationary constant-mean signal with a given IQR would be similar to the fencing with the ``exact" quartile filters in a moving boxcar window~${B_{\Delta{T}}(t) = \left[ \theta(t) - \theta(t\!-\!\Delta{T}) \right]/\Delta{T}}$, where~${\Delta{T}=2\times{\rm IQR}/\mu}$ and~$\mu$ is the QTF rate parameter.

Even though robustness of the fences generally increases as the rate parameter~$\mu$ becomes smaller, inversely, their overall tightness decreases with~$\mu$. As discussed in Section~\ref{subsubsec:less tight yet inclusive}, in order for an outlier in~$x(t)$ to protrude from the range, both its magnitude and its slew rate at a fence need to be sufficiently large. Thus the overall effectiveness of the QTF fencing for identification and mitigation of outliers is achieved through a compromise between the tightness and robustness of the fences, as illustrated in Fig.~\ref{fig:QTFs4noise}.

\begin{figure}[!t]
\centering{\includegraphics[width=8.6cm]{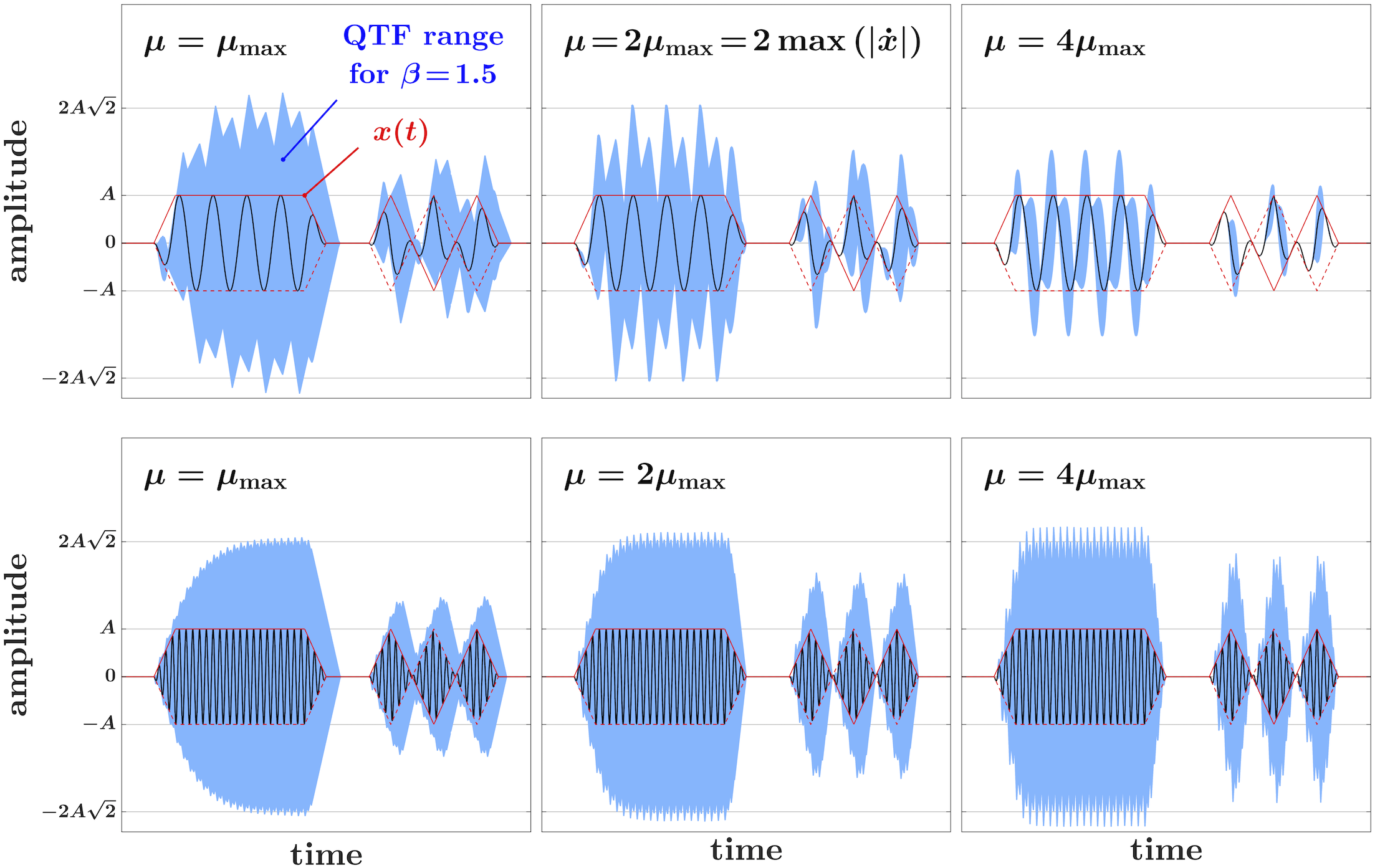}}
\caption{QTF fences for amplitude-modulated signals.
\label{fig:modulated}}
\end{figure}
\begin{figure*}[!b]
\centering{\includegraphics[width=16cm]{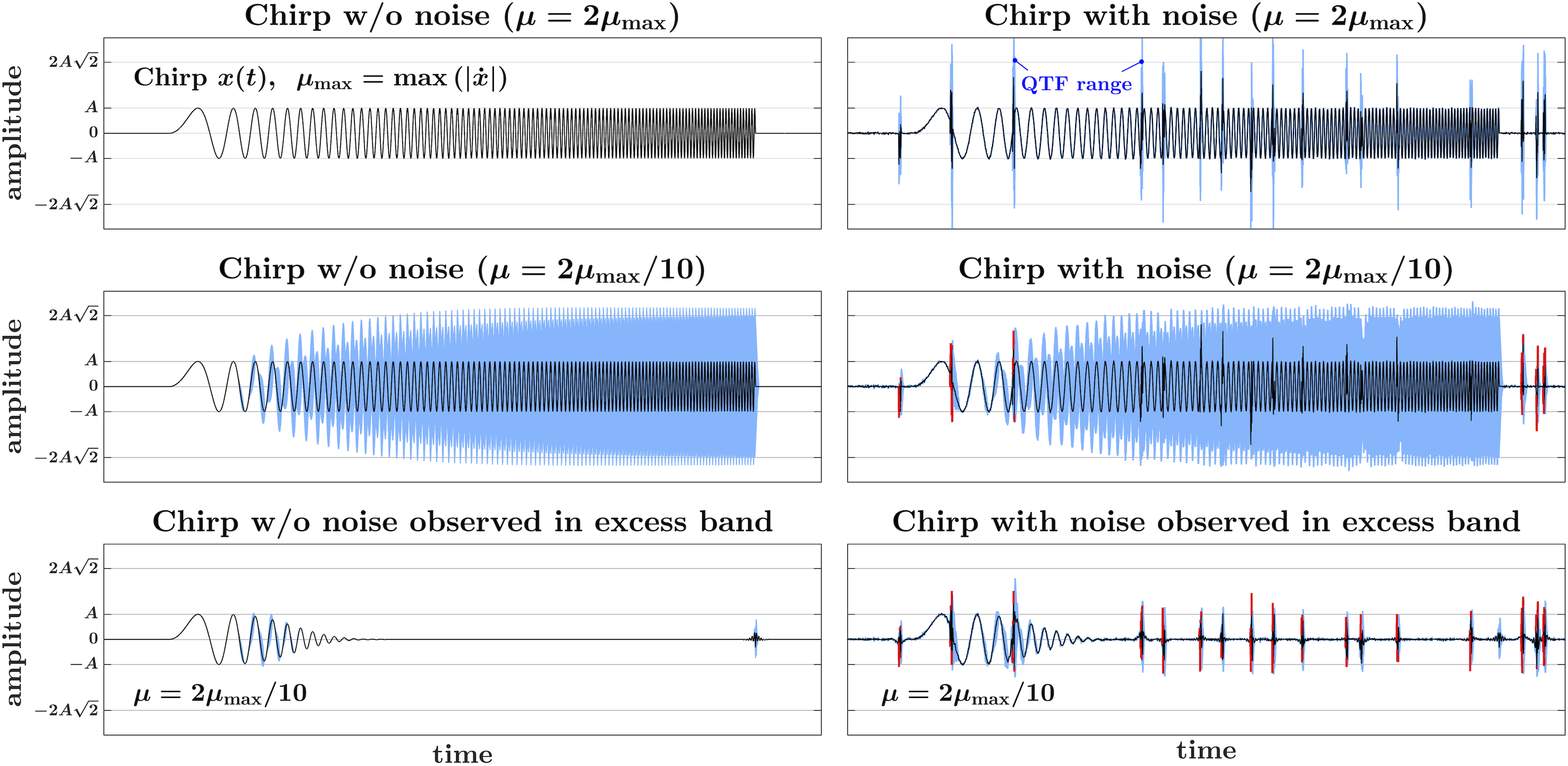}}
\caption{Suppressing high-frequency portion of chirp signal with excess band filter shown in Fig.~\ref{fig:QTFs4chirp bands} enables more reliable identification of wideband outlier noise by increasing tightness and robustness of QTF fences.
{\em Red lines: Signal+noise protruding outside of QTF fences.}
\label{fig:QTFs4chirp}}
\end{figure*}

\subsection{Fences for slowly varying signals affected by wideband noise} \label{subsec:slowly varying}
A typical task would be constructing tight fences for a slowly varying signal affected by a wideband noise. For this, as follows from the discussion in~Sections~\ref{subsubsec:linear trend} and~\ref{subsubsec:tightest}, the rate parameter~$\mu$ of the QTFs for~$Q_{[1]}(t)$ and~$Q_{[3]}(t)$ must be larger than twice the maximum slew rate of the signal of interest, to ensure that the mid-range does not diverge from the signal. With this constraint, as follows from~(\ref{eq:range protrusion}), the QTF fencing can only identify outliers of sufficiently large magnitude and the slew rate at a fence larger than~$3\mu$.

Let us assume that an outlier is produced by the response of a filter to a short-duration event, e.g., an impulse or a step. For a given filter type, the magnitude of this response will be proportional to the bandwidth of the filter, and the slew rate of the response would be generally proportional to the square of the filter bandwidth. Thus the QTF fencing would be generally more effective for identification and mitigation of outliers in a wideband noise affecting a slowly varying signal at higher observation bandwidths. This is illustrated in Fig.~\ref{fig:QTFs4ramp} for several mixtures of impulsive and Gaussian noises observed at two different bandwidths, $\Delta{f}$~and~$4\Delta{f}$. As can be seen in the figure, at a lower bandwidth the impulsive noise is not identified. In contrast, at a wider bandwidth the impulsive noise appears as distinct outliers in the noise mixture, and these outliers are identified by the QTF fencing.

\subsection{QTF fences for amplitude-modulated signals} \label{subsec:amplitude modulated}
For an amplitude-modulated signal ${x(t)\sin\left(2\pi f_{\rm c}t\right)}$ with ${\max \left( \left|\dot{x}(t)\right| \right) = \mu_{\rm max}}$, the condition for the QTF fences with~${\beta=1.5}$ to be inclusive of such a signal for any carrier frequency~$f_{\rm c}$ is~${\mu\ge\mu_{\rm max}}$, as illustrated in Fig.~\ref{fig:modulated} for low (upper panels) and high (lower panels)~$f_{\rm c}$. Note that for a given~$\mu$ and a constant modulating signal ${x(t)=A}$ the QTF outputs $Q_{[3]}(t)$ and $Q_{[1]}(t)$ converge to $A/\sqrt{2}$ and $-A/\sqrt{2}$, respectively. When ${f_{\rm c}\gg \mu/A}$, and for ${\beta=1.5}$, the QTF range becomes~$[-2A\sqrt{2},2A\sqrt{2}]$. For such a range, even for maximally robust fences (i.e. in the limit ${\mu\to 0}$) the necessary condition for reliable identification of additive outliers will be that their magnitudes exceed ${(2\sqrt{2}+1)A\approx 3.83A}$, and the outliers with the magnitudes smaller than~${(2\sqrt{2}-1)A\approx 1.83A}$ cannot be identified at all. 

\section{Increasing tightness and robustness of QTF fences by complementary filtering} \label{sec:increasing tightness and robustness}
As discussed in the previous section, a compromise between the tightness and robustness of the QTF fences can be more difficult to achieve for the signals with a strong high-frequency component, especially when this component has a large PAR and, therefore, a larger maximum slew rate compared to smaller-PAR signals with the same PSD. For example, waveforms due to orthogonal frequency-division multiplexing (OFDM) would typically have large PARs and be particularly challenging for tight yet robust QTF fencing~\cite{Nikitin19complementary}. The {\em Complementary\/} Intermittently Nonlinear Filtering (CINF) introduced in~\cite{Nikitin19hidden, Nikitin19complementary} addresses this challenge.

\begin{figure}[!b]
\centering{\includegraphics[width=8.6cm]{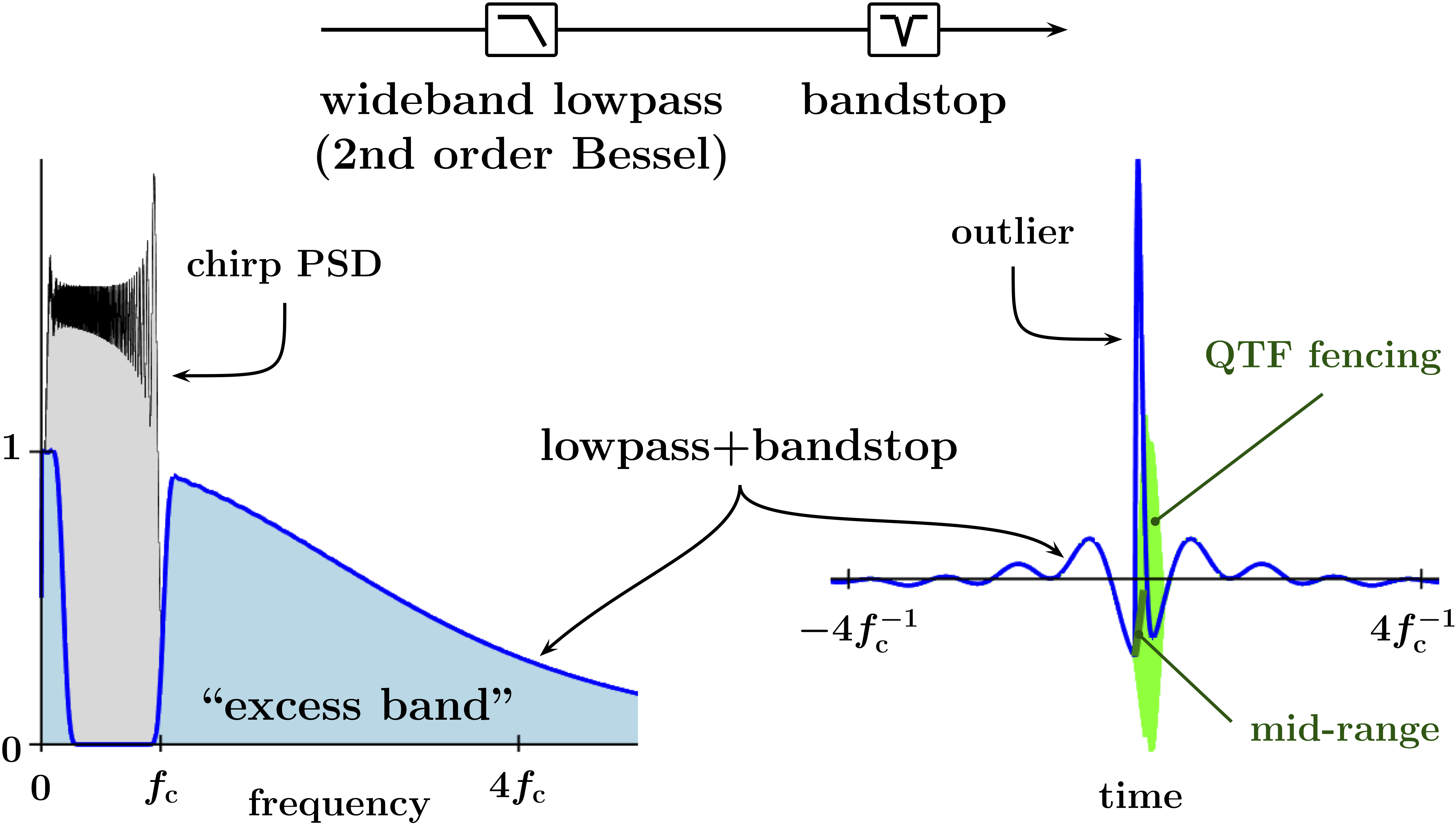}}
\caption{Impulse and frequency responses of excess band filter for chirp.
\label{fig:QTFs4chirp bands}}
\end{figure}

\subsection{QTF fences for chirps} \label{subsec:chirps}
Let us first consider the signal of interest~$x(t)$ that is a wideband linear up-chirp. For a given chirp amplitude, ${\mu_{\rm max}=\max\left(\left|\dot{x}(t)\right|\right)}$ is determined by the highest frequency in the chirp, and for the tightest possible fences the QTF rate parameter must be at least~${\mu=2\mu_{\rm max}}$ (see Section~\ref{subsubsec:tightest}). However, as shown in Section~\ref{subsubsec:less tight yet inclusive}, the QTF range will remain fully inclusive unless the slew rate of the signal+noise mixture exceeds~$6\mu$. Hence such fencing will not be robust to small-amplitude outliers, as illustrated in the upper panels of Fig.~\ref{fig:QTFs4chirp}.

Robustness of the fences can be significantly increased by reducing the QTF rate parameter, e.g. by an order of magnitude. However, while being inclusive of the signal, such fencing will only remain tight for the lower-frequency portion of the chirp, and, as can be seen in the middle panels of Fig.~\ref{fig:QTFs4chirp}, the noise outliers affecting the higher-frequency portion of the chirp are still not identified.

Consider the ``excess band" filter shown in Fig.~\ref{fig:QTFs4chirp bands}. In this example, the 3\,dB corner frequency of the front-end lowpass filter is~$3f_{\rm c}$, where~$f_{\rm c}$ is the highest frequency of the chirp, and the Bessel response ensures a small time-bandwidth product of the filter. This wideband filter is cascaded with a linear phase bandstop filter, where the high-frequency edge of the stopband is at about~$f_{\rm c}$, and the low-frequency edge is placed at approximately~$f_{\rm c}/5$. Such a bandstop filter will reduce the maximum slew rate of a linear chirp by about an order of magnitude, and, as shown in the lower left panel of Fig.~\ref{fig:QTFs4chirp}, the QTF fencing with~${\mu=\mu_{\rm max}/5}$ will become tight for the chirp filtered with the excess band filter. Still, the excess band filter applied to wideband noise will mainly preserve the outlier structure of the noise. The impulse response of the excess band filter can indeed be viewed as the difference between the impulse response of the wideband Bessel filter (which is dominated by a tall narrow pulse) and the impulse response of a bandpass filter for the passband~${[f_{\rm c}/5,f_{\rm c}]}$, which will have about an order of magnitude smaller amplitude and approximately two orders of magnitude smaller maximum slew rate than the response of the Bessel filter. Thus, outliers in wideband noise affecting the chirp will mainly remain outliers after the convolution with such an impulse response, and, as shown in the lower right panel of Fig.~\ref{fig:QTFs4chirp}, the QTF fencing with the same rate parameter~${\mu=\mu_{\rm max}/5}$ will reliably identify such outliers.

\begin{figure}[!b]
\centering{\includegraphics[width=8.6cm]{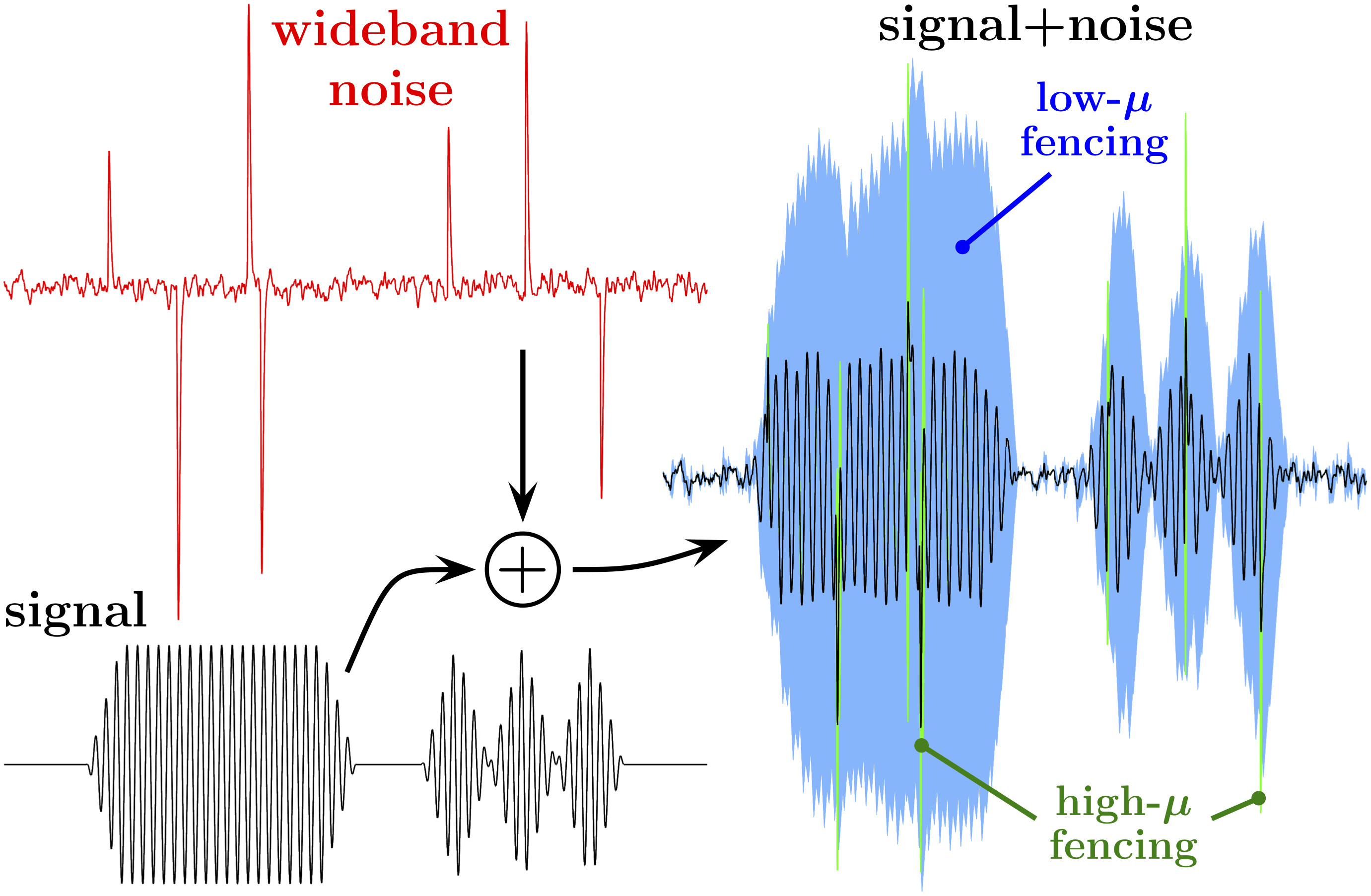}}
\caption{Signal+noise with ``obscured" noise outliers that are not identified by QTF fencing.
\label{fig:QTFs4modulation signal}}
\end{figure}
\begin{figure*}[!t]
\centering{\includegraphics[width=16cm]{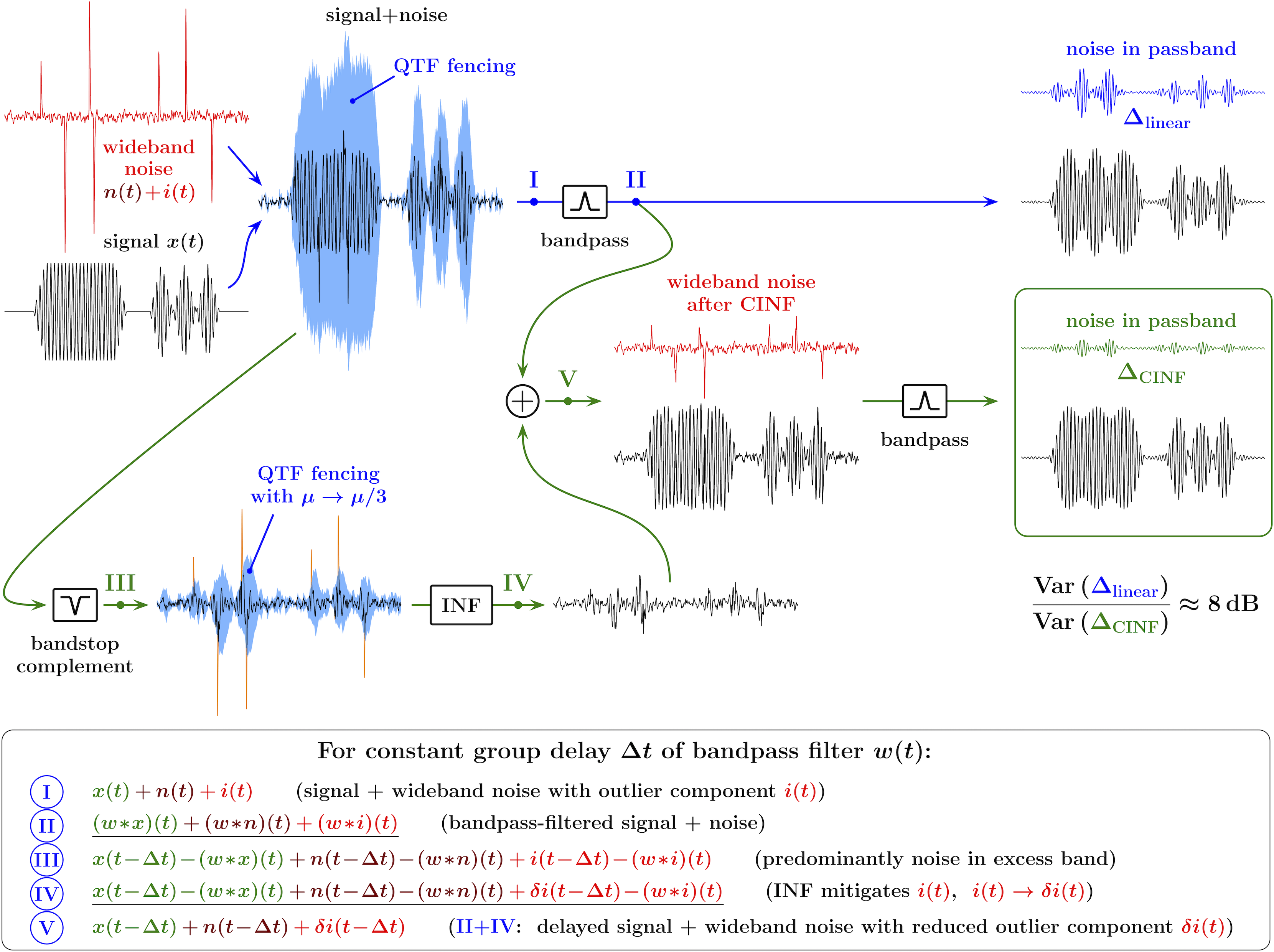}}
\caption{Complementary INF for removing wideband noise outliers while preserving band-limited signal of interest.
\label{fig:QTFs4modulation}}
\end{figure*}

\subsection{Complementary INF} \label{subsec:complementary}
As was just discussed, the impulse response of an excess band filter constructed as a bandstop filter cascaded with a wider-band, small time-bandwidth product bandpass filter will contain a distinct outlier that can be easily identified. For example, as illustrated in Fig.~\ref{fig:QTFs4chirp bands}, QTF fencing that is tight for the impulse response of the bandpass filter can still reliably identify the outlier in the impulse response of the excess band filter. Also note that replacing the outlier values  in the impulse response of the excess band filter with those in mid-range of the fencing effectively converts the excess band filter into its ``complementary" bandpass filter. Thus the removed outlier in the impulse response of the excess band filter itself approximates the response of the front-end wideband filter. This ability of INF to ``invert" an excess band filter, by removing the outlier in its impulse response, enables the construction of robust and inclusive, yet tight, QTF fences for identification and removal of wideband noise outliers from a narrower-band signal in a manner previously illustrated in Fig.~\ref{fig:INF fencing}.

Let us present a more detailed illustration of a CINF arrangement for mitigation of wideband outlier noise affecting a narrowband signal of interest, and the CINF's ability to increase both tightness and robustness of QTF fencing. For this example, we use a signal+noise mixture where the outliers in the wideband noise are additionally obscured by the amplitude-modulated signal of interest (see Fig.~\ref{fig:QTFs4modulation signal}). For this, the QTF fencing (shown by the green shading) that is tight for the signal of interest is {\bf not sufficiently robust,} as it remains inclusive for the total signal+noise mixture and does not identify noise outliers. However, the fencing (blue shading) that is more robust yet still inclusive of the signal of interest is {\bf not sufficiently tight} for identification of outliers (see Section~\ref{subsec:amplitude modulated}).

In Fig.~\ref{fig:QTFs4modulation}, the bandpass filter mainly matches the signal's passband, and the bandstop filter is its ``complement" obtained by spectral inversion of the bandpass filter, so that the sum of the outputs of the bandpass and the bandstop filters is equal to the input signal. The output of the bandpass filter applied to the signal+noise mixture is shown in the upper right corner of the figure, where the trace marked by ``$\Delta_{\rm linear}$" shows the passband noise.

Since the bandstop filter ``blocks" the signal of interest, its output is mainly the ``excess band" noise, which is the difference between the ``original" wideband noise and the noise filtered with the bandpass filter. As discussed earlier, the outliers in the excess band noise can be reliably identified by QTF fencing. In addition, an effective compromise between the tightness and robustness of the fences is more easily achieved for the excess band noise only, rather than for the original signal+noise mixture. For example, the rate parameter for QTF fencing of the excess band noise in Fig.~\ref{fig:QTFs4modulation} is three times smaller than that required for the fencing of the signal+noise mixture to remain inclusive of the signal. After the outliers in the excess band noise are mitigated by the INF, the remaining excess band noise is added to the output of the bandpass filter. This combined output will be equal to the original signal of interest affected by a wideband noise with reduced outliers. As the result, after bandpass filtering of this signal+noise mixture with reduced noise outliers, the passband noise is significantly reduced. This is shown in the lower right part of the figure, where the trace marked by ``$\Delta_{\rm CINF}$" shows the passband noise after CINF.

\begin{figure}[!t]
\centering{\includegraphics[width=8.6cm]{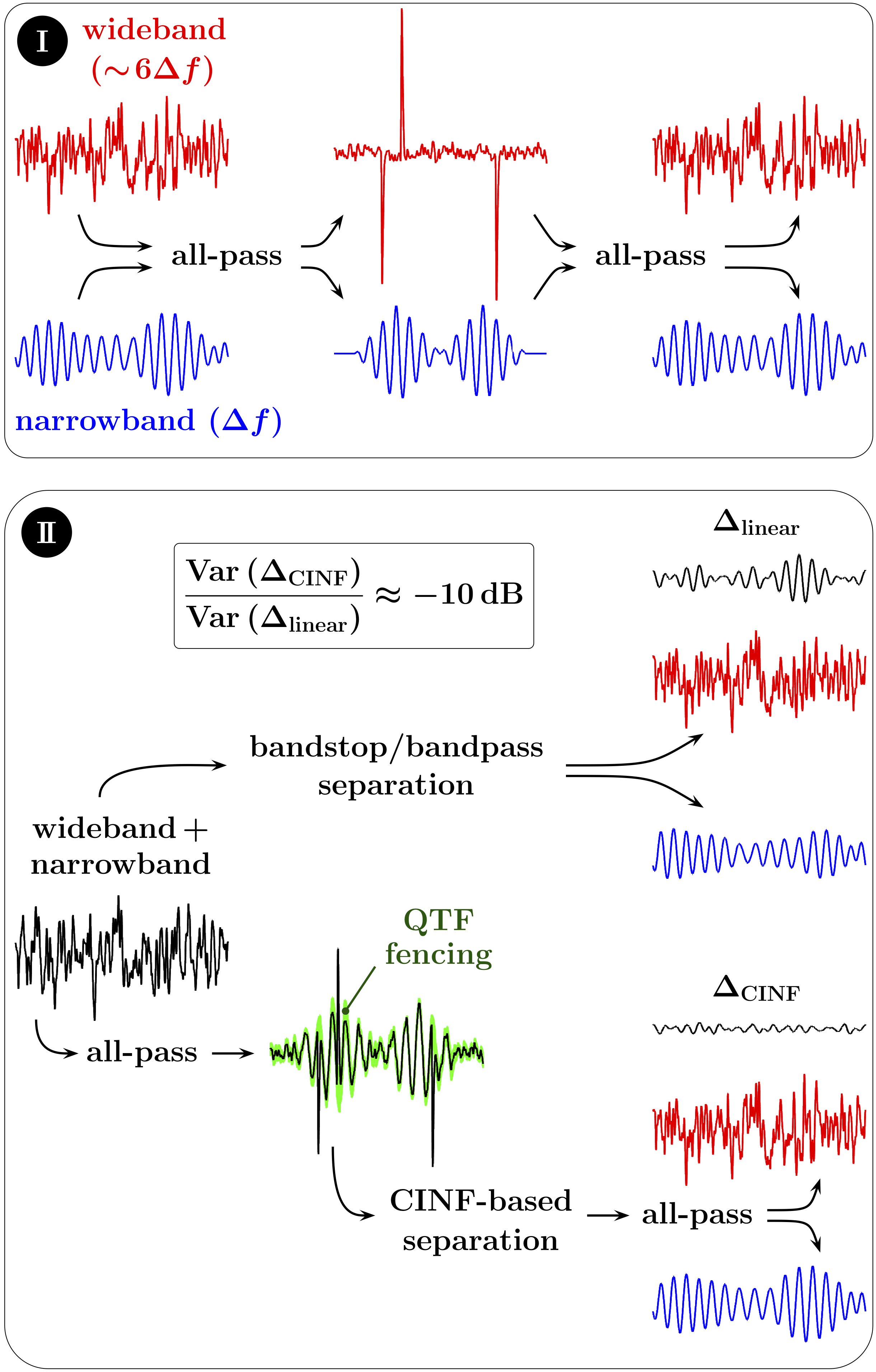}}
\caption{CINF-based separation of wideband and narrowband signals.
\label{fig:QTFs wide narrow}}
\end{figure}

\section{Conclusion} \label{sec:conclusion}
The main focus of this paper is the basic properties, parameters, and behavior of robust fences for Intermittently Nonlinear Filtering (INF), established by Quantile Tracking Filters (QTFs). QTFs are an appealing choice for such robust fencing in INF due to their simplicity, efficiency, and effectiveness. They are conceptually {\em analog\/} filters suitable for real-time processing of continuous-time signals and are easily implementable in analog circuitry. Moreover, their numerical computations are {\boldmath$\mathcal{O}(1)$} per output value in both time and storage, which also enables their high-rate digital implementations in real, or near real, time. Along with the QTF fencing, we also discuss the INF and the complementary INF arrangements in general, expounding on many details that were either abbreviated or omitted in our previous publications on INF.

While the main use of the INF discussed so far has been the mitigation of wideband noise affecting a narrowband signal, it can just as easily be applied to mitigation of narrowband interference with a wideband signal. For example, it can be used for enabling narrowband interference rejection in spread-spectrum communications in excess of that provided by the process gain only.

More broadly, the INF can be used for enhanced separation of different waveforms with overlapping spectral content in general, facilitating synthesis of dual function systems and hard-to-intercept communications links. Fig.~\ref{fig:QTFs wide narrow} provides a simplified example of such separation for a wideband and a narrowband signals. As discussed in Section~\ref{sec:introduction}, apparent outliers in a signal can appear, disappear, and reappear due to linear filtering, and the effect of filtering on the temporal and/or the amplitude structure of a signal would be more apparent at a wider bandwidth. In a wide band, such filtering can drastically change the time-domain appearance of the signal and its amplitude density. In a narrow band, these changes would be much less apparent. This is illustrated in panel~I of Fig.~\ref{fig:QTFs wide narrow}, where an all-pass filter creates distinct outliers in the wideband signal, while having much less impact on the narrowband signal. Note that a conjugate all-pass filter can then restore the original time-domain appearance of both signals.

As shown in panel~II of Fig.~\ref{fig:QTFs wide narrow}, the mixture of the two signals after the all-pass filtering can be treated as a wideband outlier noise affecting a narrowband signal of interest. Then these signals can be effectively separated by the CINF, and the conjugate all-pass filter can be used to restore their original time-domain appearances. Since the ratio of the bandwidths is only about~8\,dB, the error in the separation of these signals achievable by a linear complementary bandstop/bandpass pair (indicated as $\Delta_{\rm linear}$) is rather large. In contrast, the error of the CINF-based separation ($\Delta_{\rm CINF}$) is significantly smaller.

\appendices
\section{Numerical implementation of QTFs} \label{app:numerical}
Since outputs of analog QTFs are piecewise-linear signals consisting of alternating segments with positive and negative slopes, care should be taken in finite difference implementations of QTFs to avoid the ``overshoots" around the crossings of~$Q_q(t)$ with~$x(t)$. In particular, when $x(n)\!-\!Q_q(n-1)$ is outside of the interval ${h\mu\left[2(q-1),2q\right]}$, where~$h$ is the time step, one may set $Q_q(n)\!=\!x(n)$, as illustrated in~\cite{Nikitin19ADiCpatentCIP1} and in the MATLAB function below:\\

\protect\renewcommand{\baselinestretch}{1} \footnotesize
\noindent
\begin{minipage}{\textwidth}
\begin{alltt}
%---------------------------------------------------
\textbf{function y = QTFs(x,dt,mu,q)}
%---------------------------------------------------
\textbf{lx = length(x);  lq = length(q);}
\textbf{q = q(:);  y = zeros(lx,lq);  gamma = mu*dt;}
\textbf{y(1,:) = x(1)*ones(1,lq);}
%---------------------------------------------------
\textbf{for i = 2:lx}
\textbf{  dX = x(i)*ones(1,lq)-y(i-1,:);}
\textbf{  for j = 1:lq}
\textbf{    if dX(j)>2*gamma*(q(j)-1) & dX(j)<2*gamma*q(j)}
\textbf{      y(i,j) = x(i);}
\textbf{    else}
\textbf{      y(i,j)=y(i-1,j)+gamma*(sign(dX(j))+2*q(j)-1);}
\textbf{    end}
\textbf{  end}
\textbf{end}
\textbf{return}
%---------------------------------------------------
\end{alltt}
\end{minipage}\\
\protect\renewcommand{\baselinestretch}{1} \small\normalsize

\section*{Acknowledgment}
The authors would like to thank
Kendall Castor-Perry (aka The Filter Wizard), and
Kyle~D. Tidball of Textron Aviation, Wichita, KS,
for heir valuable suggestions and critical comments.
This work was supported in part by Pizzi Inc., Denton, TX 76205 USA.


\end{document}